\documentclass[a4paper,11pt]{article}
\pdfoutput=1 
\usepackage{jheppub} 

\usepackage{array}
\usepackage{comment}
\usepackage{braket}
\usepackage{cancel}
\usepackage{physics}
\usepackage{slashed}
\usepackage{textcase}
\usepackage{graphicx,epsfig,psfrag,amssymb,hyperref,cleveref}
\usepackage{multirow, multicol}
\usepackage{xcolor,graphicx,epsfig,psfrag,amsmath,empheq}
\usepackage{bm}
\usepackage{mathrsfs,amsfonts,soul,color}
\usepackage{xspace}
\usepackage[caption=false]{subfig}
\usepackage{hepunits}
\counterwithin*{equation}{section}
\allowdisplaybreaks

\newcolumntype{?}{!{\vrule width 2pt}}

\usepackage[colorinlistoftodos, disable]{todonotes}

\newcommand{\tev}{\ensuremath{\mathrm{\: Te\kern -0.1em V}}\xspace}
\newcommand{\gev}{\ensuremath{\mathrm{\: Ge\kern -0.1em V}}\xspace}
\newcommand{\mev}{\ensuremath{\mathrm{\: Me\kern -0.1em V}}\xspace}
\newcommand{\jpc}{\ensuremath{\mathrm{J}^{\mathrm{PC}}}\xspace}

\def\BR{\ensuremath{\mathcal{B}}\xspace}
\def\aprime{\ensuremath{A^{\prime}}\xspace}
\newcommand{\deriv}[1]{\ensuremath{\mathrm{d#1}}}

\newcommand{\cL}{\mathcal{L}}
\newcommand{\cF}{\mathcal{F}}
\newcommand{\cR}{\mathcal{R}}
\newcommand{\cO}{\mathcal{O}}
\newcommand{\cC}{\mathcal{C}}

\newcommand{\eg}{\textit{e.g.}}

\newcommand{\SUF}{\mathrm{SU(3)}_{\mathrm{flavor}}}

\newcommand{\darkcast}{\textsc{DarkCast}\xspace}

\definecolor{red1}{cmyk}{0,1,1,0.3}

\begin{document}
\listoftodos

\title{Axial vectors in DarkCast}

\author[a]{Chaja Baruch}
\author[b]{Philip Ilten}
\author[a]{Yotam Soreq}
\author[c]{Mike Williams}

\affiliation[a]{Physics Department, Technion—Israel Institute of Technology, Haifa 3200003, Israel}
\affiliation[b]{Department of Physics, University of Cincinnati, Cincinnati, OH, USA}
\affiliation[c]{Laboratory for Nuclear Science, Massachusetts Institute of Technology, Cambridge, MA, USA}

\emailAdd{
chajabaruch@campus.technion.ac.il, 
philten@cern.ch, 
soreqy@physics.technion.ac.il, 
mwill@mit.edu}

\abstract{
In this work, we explore new spin-1 states with axial couplings to the standard model fermions. 
We develop a data-driven method to estimate their hadronic decay rates based on  data from $\tau$ decays and using SU(3)$_{\rm flavor}$ symmetry. 
We derive the current and future experimental constraints for several benchmark models. 
Our framework is generic and can be used for models with arbitrary vectorial and axial couplings to quarks.
We have made our calculations publicly available by incorporating them into the \darkcast package, see \url{https://gitlab.com/darkcast/releases}. 
}

\maketitle

\section{Introduction} 

The standard model~(SM) is an extremely successful theory of physics at the fundamental level, but it does not provide a complete description of nature, and therefore, must be extended in some way. 
Substantial effort has been dedicated in recent years to probing new physics~(NP)~\cite{EuropeanStrategyforParticlePhysicsPreparatoryGroup:2019qin}, in particular at the MeV-to-GeV mass scale, see \eg~\cite{Beacham:2019nyx}. 
Many of these efforts are searching for the \textit{dark photon}~\cite{Okun:1982xi,Galison:1983pa,Holdom:1985ag,Pospelov:2007mp,ArkaniHamed:2008qn,Bjorken:2009mm}, $\aprime$, a new massive spin-1 particle that kinetically mixes with the ordinary photon. 
Dark photon searches have been conducted by numerous experiments, 
among them beam-dump~\cite{Bergsma:1985is,Konaka:1986cb,Riordan:1987aw,Bjorken:1988as,Bross:1989mp,Davier:1989wz,Athanassopoulos:1997er,Astier:2001ck,Bjorken:2009mm,Essig:2010gu,Williams:2011qb,Blumlein:2011mv,Gninenko:2012eq,Blumlein:2013cua,Banerjee:2018vgk,NA64:2019imj}, fixed-target~\cite{Abrahamyan:2011gv,Merkel:2014avp,Merkel:2011ze,Essig:2010xa,Moreno:2013mja,Adrian:2018scb}, collider~\cite{Aubert:2009cp,Curtin:2013fra,Lees:2014xha,Ablikim:2017aab,Aaij:2017rft,Anastasi:2015qla,Anastasi:2018azp,Aaij:2019bvg,Sirunyan:2019wqq,Lees:2017lec,Abdallah:2003np,Abdallah:2008aa}, and rare-meson-decay~\cite{Bernardi:1985ny,MeijerDrees:1992kd,Archilli:2011zc,Gninenko:2011uv,Babusci:2012cr,Adlarson:2013eza,Agakishiev:2013fwl,Adare:2014mgk,Batley:2015lha,KLOE-2:2016ydq,CortinaGil:2019nuo} experiments. 
Moreover, many proposals~~\cite{Essig:2010xa,Freytsis:2009bh,Balewski:2013oza,Wojtsekhowski:2012zq,Beranek:2013yqa,Echenard:2014lma,Battaglieri:2014hga,Alekhin:2015byh,Gardner:2015wea,Ilten:2015hya,Curtin:2014cca,He:2017ord,Kozaczuk:2017per,Ilten:2016tkc,Feng:2017uoz,Craik:2022riw,Galon:2022xcl,Curtin:2018mvb,Gligorov:2017nwh,Raggi:2014zpa,Nardi:2018cxi,Seo:2020dtx,Tsai:2019mtm,Gan:2020aco,Gninenko:2019qiv,Ambrosino:2019qvz,Akimov:2019xdj,Battaglieri:2016ggd,SHiP:2020noy,Doria:2019sux,Akesson:2018vlm} for probing unexplored $\aprime$ parameter space (mass, kinetic mixing, and invisible decay rate) have been put forward.  
For recent reviews see~\cite{Graham:2021ggy,Fabbrichesi:2020wbt}. 

The dark photon searches in the MeV-to-GeV mass range can be be reinterpreted in a broader context of new feebly interacting massive particles~(FIMPs). 
In particular, they can be recast for generic vector models in the \darkcast framework~\cite{Ilten:2018crw}, see also~\cite{Bauer:2018onh}.
In addition to a vector-like coupling, a new massive spin-1 boson can couple to the axial current of the SM fermions. 
For example, models with a non-vanishing axial coupling and a mass below that of the pion were explored in~\cite{Kahn:2016vjr}.

One major challenge in probing new physics at the MeV-to-GeV mass scale with couplings to quarks and/or gluons is reliably estimating the hadronic decay rates of the new states. 
Many current and near future experiments have potential sensitivity to new physics at this mass scale; thus, it is important to reliably estimate these hadronic rates. 
For sub-GeV masses, chiral perturbation theory can be used in several cases, such as for pseudo-scalars.
Above several GeV, perturbative QCD holds and can be utilized to calculate inclusive rates. 
However, the mass range between these two regimes is challenging.

One possible avenue for dealing with the region where neither chiral perturbation theory nor perturbative QCD is valid is developing data-driven methods. 
The $e^+e^-\!\!\to$\,hadrons data along with $\SUF$ symmetry is of great use in this regard. 
Two successful examples are determining the hadronic rates of new spin-1 bosons with vectorial coupling, see \darkcast~\cite{Ilten:2018crw} and \cite{Foguel:2022ppx} for recent progress; 
and determining the hadronic rates of pseudo-scalars~\cite{Aloni:2018vki}, see also~\cite{Cheng:2021kjg}, where additionally crossing symmetry plays an important role.

In this work, we study the scenario of new massive spin-1 bosons with chiral couplings (axial and vectorial) at the intensity frontier.
In particular, we develop a data-driven method to estimate their hadronic decay rates based on data from $\tau$ decays and using $\SUF$ symmetry. 
We recast existing experimental results into constraints on the parameter space for several benchmark models.
Our method can be systematically applied to any spin-1 model with couplings to the SM fermions (quarks and leptons).
In addition, the predictions obtained using our framework can be improved by incorporating future higher-precision data.
Finally, we include our results as part of the \darkcast framework, see \url{https://gitlab.com/darkcast/releases}. 

The rest of this article is organized as follows. 
\Cref{sec:model} introduces the generic model for a spin-1 particle with both vector and axial-vector couplings to the SM fermions. 
It provides the means for recasting dark-photon bounds to a model with purely axial couplings. 
This includes a data-driven method of obtaining the hadronic decay widths. 
\Cref{sec:experiments} discusses the experiments that provide the dark-photon limits we recast. 
We provide three examples for the application of our framework in \cref{sec:examples}: a purely axial boson, a boson with chiral couplings, {\em i.e.}\ both nonzero axial and vector couplings, and a 2-Higgs-doublet model. 
\Cref{sec:summary} provides a summary and some concluding remarks. 

\section{Generic Chiral Boson Model}
\label{sec:model}

We consider a generic model with a spin-1 boson, $X$, that has both vector and axial-vector couplings to the SM fermions, $f$, as well as couplings to dark sector states, $\chi$, that we do not specify.
The effective $X$ interactions can be written as
\begin{align}
    \cL 
=   g_X\sum_f \bar{f} \left(x^f_{V}\gamma^\mu 
    + x^f_{A}\gamma^\mu \gamma^5\right)f X_{\mu} 
    + \sum_\chi \cL_{X_{\chi\bar{\chi}}}\,,
\end{align}
where $g_X x^f_{V,A}$ is the strength of the interaction between $X_\mu$ and the axial $(A)$ or vector~$(V)$ currents of the SM fermions.  
The canonical dark photon model~\cite{Holdom:1985ag}, where $X \equiv \aprime$, is given by $g_X=e\epsilon$, with $\varepsilon$ the kinetic mixing parameter, $x_V^\ell=-1$, $x_V^\nu=0$, $x_V^{d,s,b}=-1/3$, $x_V^{u,c,t}=+2/3$, and all $x^f_{A}=0$. 
(Note that in this work, we consider only flavor-diagonal and CP-conserving interactions.)

Many existing experimental constraints have been placed on the \aprime model. 
To recast a dark-photon search that used the final state $\cF$, we solve
\begin{align}
    \sigma_X \BR_{X\to\cF} \epsilon\left(\tau_X\right) 
=   \sigma_{\aprime} \BR_{\aprime\to\cF} \epsilon\left(\tau_{\aprime}\right)
\end{align}
at each $m_{\aprime}=m_X$, where $\sigma_{X,\aprime}$ are the $X$ and $\aprime$ production cross sections, $\BR_{X,\aprime \to \cF}$ denotes the decay branching fractions to the final-state $\cF$, and $\epsilon(\tau)$ is the lifetime-dependent detector efficiency. 
The production and decay ratios, $\sigma_X / \sigma_{\aprime}$ and $\BR_{X\to\cF} / \BR_{\aprime\to\cF}$, are determined in the following two subsections. 
We use the same approximations for the efficiency ratio, $\epsilon\left(\tau_X\right) / \epsilon\left(\tau_\aprime\right)$, as in \darkcast~\cite{Ilten:2018crw}.
Since the $X$ or $\aprime$ would be highly boosted in these experiments, the differences in the angular acceptance between the decays of vector and chiral bosons will be small and are neglected. 

In the next two subsections, we study the production and decay ratios of a purely axial boson. 
For chiral models, where both vector and axial-vector couplings are present, we have analytically confirmed that there is no interference between the vector and axial currents in leptonic production and decay, as well as for quarks in the perturbative region. 
In addition, we checked that for the decay of a chiral boson into two- and three-meson final states there is no interference between the axial and vector currents. 
This makes recasting straightforward: for any final state that can be reached by both a vector and axial current, the total cross section or decay width is just the sum of the vector and axial components. 
However, if a case is found where vector-axial interference is required, it is straightforward to include such contributions.  
The rest of this section focuses on the purely axial case, since the purely vector case was already studied in \cite{Ilten:2018crw}. 

\subsection{Production of a purely axial boson}
\label{sec:prod}

We now determine the production cross section ratios between a purely axial vector boson ($x^f_{A}\ne0$ and $x^f_{V}=0$) and the dark photon, for the following dark-photon production mechanisms: electron and proton bremsstrahlung, $e^+e^-$ annihilation, Drell-Yan production, and several important meson decays. 

The production cross sections for electron bremsstrahlung and $e^+e^-$ annihilation are the same as for the \aprime, modulo the fermion coupling strengths, up to a correction of $\cO (m_e^2/m_X^2)$: 
\begin{align}
    \frac{\sigma(e^+e^-\to \gamma X)}{\sigma(e^+e^-\to \gamma \aprime)} 
=   \frac{\sigma(eZ\to eZX)}{\sigma(eZ\to eZ\aprime)} 
=   \frac{(g_X x^e_{A})^2}{\varepsilon^2 e^2} 
    \left[ 1+ \cO \left(\frac{m_e^2}{m_X^2}\right)\right]\,.
    \label{eq:Abrem}
\end{align} 
The A1~\cite{Merkel:2014avp} and APEX~\cite{Abrahamyan:2011gv} experiments, the NA64 experiment~\cite{Banerjee:2019pds} as well as  the E141, E137, E774, KEK, and Orsay electron beam-dump experiments \cite{Riordan:1987aw,Bjorken:1988as,Bross:1989mp,Konaka:1986cb,Davier:1989wz}, all searched for a dark photon produced through electron bremsstrahlung.
In addition, the NA64$_\mu$ experiment~\cite{Gninenko:2019qiv} will search for a dark photon produced via muon bremsstrahlung.
Recasting this future bound is straightforward, but the full expression for bremsstrahlung production must be taken into account. 

For proton bremsstrahlung, which is used by the  $\nu$-CAL~I~\cite{Blumlein:1990ay,Blumlein:1991xh, Blumlein:2013cua} experiment, we can to a good approximation take the axial charge to be $2x^u_{A}+x^d_{A}$, which gives
\begin{align}
    \frac{\sigma(pZ\to pZX)}{\sigma(pZ\to pZ\aprime)} 
=   \frac{g_X^2 \left(2x^u_A+x^d_A\right)^2}{\varepsilon^2 e^2} 
    \left(\frac{F_A(m_X)}{F_V(m_X)}\right)^2 \,,
\end{align}
where the ratio of the form factors of the proton is~\cite{Bodek:2007ym} 
\begin{align} 
        \left(\frac{F_A(m_X)}{F_V(m_X)}\right)^2
\approx 1.6 \left(\frac{1+\left(\frac{m_X}{1.01 \gev}\right)^2}
        {1+\left(\frac{m_X}{0.84 \gev}\right)^2}\right)^4\,.
\end{align}
These form factors are obtained within the dipole approximation, 
which is approximately valid in the mass range in which proton bremsstrahlung is an important production mechanism.

For Drell-Yan~(DY) production, which is relevant for the LHCb dark-photon searches \cite{Ilten:2016tkc,Aaij:2017rft},  
we can write the ratio of dark-photon and axial-boson cross sections as a sum over quark flavors as follows:
\begin{align}
    \label{eq:DY}
    \frac{\sigma(\text{DY}\to X)}{\sigma(\text{DY}\to \aprime)} 
=   \sum_{q_i} \frac{\sigma(q_i\bar{q_i}\to \gamma^\ast)}{\sigma(\text{DY}\to\gamma^\ast)}
    \frac{\sigma(q_i\bar{q_i}\to X)}{\sigma(q_i\bar{q_i}\to \aprime)}\,,
\end{align}
where the first term is the fraction of the DY production attributed to each flavor in the SM, and the second term is the contribution from each sub-process. 
The contribution $\sigma(q_i\bar{q_i}\to X)$ is calculated perturbatively 
\begin{align}
\label{eq:DYratio}
    \frac{\sigma(q\bar{q}\to X)}{\sigma(q\bar{q}\to \aprime)} 
=   \frac{(g_X x^q_{A})^2}{e^2\varepsilon^2 Q^2_{q} }
    \left[ 1+ \cO \left(\frac{m_q^2}{m_X^2}\right) \right]\,,
\end{align}
where $Q_q$ is the SM charge of the quark of flavor $i$. 
To know the fraction of DY production attributed to each flavor, the parton distribution functions must be used. 
These fractions for the LHCb search can be found in Fig.~11 of \cite{Ilten:2018crw}.

Finally, we consider $X$ production in meson decays. 
This production mechanism is used in the LHCb searches~\cite{Aaij:2017rft} below a \gev, where  $\rho,\omega,\phi\to\aprime$ as well as $\eta \to \aprime \gamma, \omega \to \aprime \pi^0$ are all important. 
In addition, meson decays were used by the KLOE experiment, $\phi \to \aprime \eta$~\cite{Archilli:2011zc}, and the NA48/2 experiment, which searched for $\pi^0\to\aprime\gamma$~\cite{Batley:2015lha}.
For a purely axial boson, there are no contributions from $\rho,\omega,\phi$ mixing as the $X$ has different quantum numbers. (Instead, there would be mixing with axial-vector mesons, {\em e.g.}\ the $f_1$; however, these mesons are not considered in dark-photon limits, and no dedicated studies of such mesons would produce competitive constraints on $X$ bosons. Hence, we ignore this production mechanism.) Using the phenomenological Lagrangian of~\cite{Roca:2003uk}, which describes the C- and P-conserving interactions between the SU(3) vector, axial vector, and pseudoscalar nonets, we find that there is no vertex contributing to $\phi \to X \eta$ or $\omega \to X \pi^0$. 
Furthermore, the $\pi^0\to X\gamma$ and $\eta \to X \gamma$ processes have no contribution from the axial anomaly~\cite{Feng:2016ysn} and non-anomalous contributions vanish according to the Sutherland-Veltman theorem~\cite{Sutherland:1967vf}. 
Therefore, a purely axial boson is not constrained by NA48, KLOE, and LHCb bounds obtained from meson decays.

\subsection{Decays of a purely axial boson}
\label{sec:decay}

The $X$ boson is assumed to decay predominantly into invisible dark-sector final states if kinematically allowed, and into SM final states otherwise.
The partial width of the $X$ boson into fermions is given by
\begin{align}
    \label{eq:Xpertdecaywidth}
    \Gamma_{X\to f\bar{f}} 
=   \frac{\cC_f \left(g_X x^{f}_A\right)^2}{12\pi} m_X 
    \left(1-4\frac{m_{f}^2}{m_X^2}\right) 
    \sqrt{1-4\frac{m_{f}^2}{m_X^2}}\, ,
\end{align}
where $\cC_f=1$ for charged leptonic decays ($\ell^+\ell^-$), $\cC_f=1/2$ for neutrinos, and $\cC_f=3$ for decays to quarks ($q\bar{q}$). 
For $X$ bosons with a mass below $2\,\gev$, the perturbative calculation fails and no longer reliably describes decays to hadrons. 
To be able to recast bounds in this regime, we adopt a data-driven approach. 
Since the \aprime couples to the electromagnetic current, its decay width into hadrons is given by
\begin{align}
    \label{eq:gamma_hadrons}
    \Gamma_{\aprime\to \mathrm{hadrons}} 
=   \Gamma_{\aprime \to \mu^+\mu^-} \cR_\mu (m_{\aprime})\,,
\end{align}
where $\cR_\mu \equiv \sigma(e^+e^-\to\mathrm{hadrons}) / \sigma(e^+e^-\to\mu^+\mu^-)$ is known experimentally, see \eg~\cite{ParticleDataGroup:2020ssz}. 
Therefore, by isolating the contributions of hadronic currents with different $\SUF$ quantum numbers, the hadronic rates of an $X$ boson with vector couplings can be estimated~\cite{Ilten:2018crw}.
Exclusive \aprime hadronic rates can be estimated by a similar relation to \cref{eq:gamma_hadrons} for the relevant final state.
Finally, see~\cite{Foguel:2022ppx} for a more recent analysis. 

For axial-vector bosons, a similar relation as \cref{eq:gamma_hadrons} can be constructed even though the reaction $e^+e^-\to \mathrm{hadrons}$ via the axial current cannot be directly measured. 
First, we note that charged axial currents are accessible via weak hadronic $\tau$ decays.  
We use the hadronic $\tau$ spectral function, along with 
$\SUF$ symmetry, to obtain the neutral axial currents needed to estimate hadronic $X$ decay rates as follows: 
\begin{align}
    \label{eq:tau decay to scattering}
    \sigma\left(e^+e^-\to \cF^0\right) \approx \frac{(g_X^2 x^e_A)^2}{4\pi s} a_1^{(s)}(s)_{\cF^-}\, ,
\end{align}
where $\cF^{0,-}$ denote exclusive neutral and charged hadronic final states that belong to the same $\SUF$ multiplet, and  
$a_1^{(s)}(s)$ is the spectral function of the (strange) axial hadronic $\tau$ decay. 
The spectral functions provide the charged $\bar{u}d$ and $\bar{u}s$ currents, which we rotate into the neutral $u\bar{u}-d\bar{d}$ and $u\bar{u}-s\bar{s}$ currents using $\SUF$ symmetry. 
For convenience, we construct a linear combination to work in the basis of isovector, isoscalar, and strange currents.

\begin{figure}[t]
    \centering
    \includegraphics[width=0.91\textwidth]{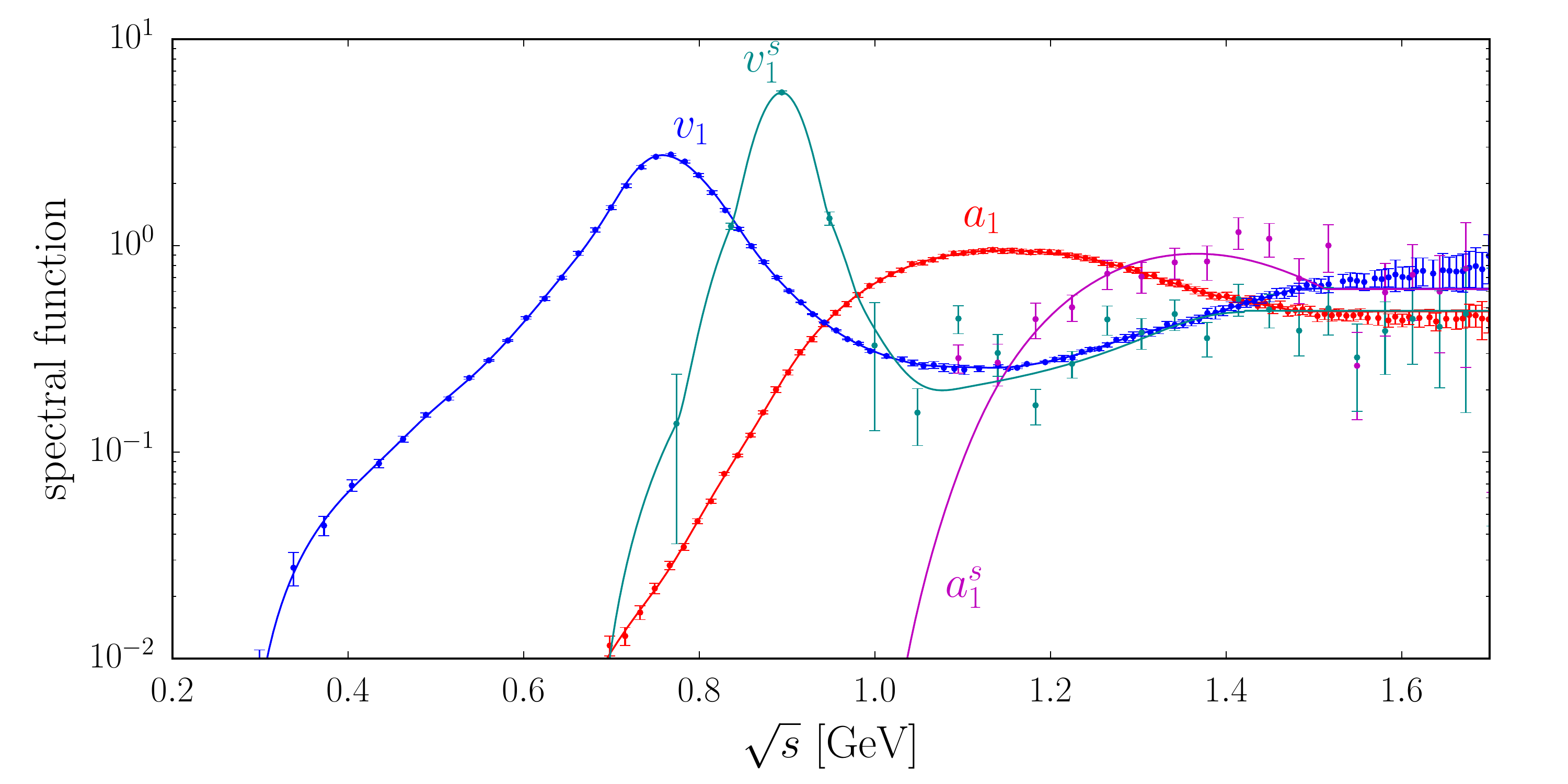}
    \caption{Data taken from the  ALEPH LEP collaboration result~\cite{ALEPH:2005qgp} used to determine the hadronic spectral $\tau$ decay functions. The lines are determined with interpolating splines and matched to their perturbative values at high center-of-mass energies. See the text and \cref{app:spectral} for further details.}
    \label{fig:spectral}
\end{figure}

The spectral functions for the axial hadronic $\tau$ decay are taken from the ALEPH LEP collaboration result~\cite{ALEPH:2005qgp}. All the spectral functions are smoothed using interpolating splines, and are matched to their perturbative values at suitably high center-of-mass energies $\sim 1.5\,\gev$. These values are $0.62$ and $0.48$ for the axial and vector strange spectral functions, respectively, and $0.48$ and $0.62$ for the axial and vector light spectral functions, respectively. See \cref{app:spectral} for further details. The spectral function fits and corresponding data are shown in \cref{fig:spectral}.
The total hadronic rate, without the sub-leading contribution due to flavor singlet states (see below), is given by (for a detailed derivation see \cref{app:hadronicwidth})
\begin{align}
    \label{eq:hadronic master}
    \Gamma_{X \to \mathrm{hadrons}} 
=  \frac{g_X^2 m_X}{4\pi} \Bigg[&(x^u_A-x^d_A)^2 a_1(m_X^2) 
    + (x^s_A)^2 \,      \Theta(m_X^2-4m_K^2) \nonumber\\
&\times\left(\frac{1}{4} a_1(m_X^2) + a_1^s(m_X^2) - \cos(\phi) \sqrt{a_1(m_X^2)a_1^s(m_X^2)}\right)\Bigg] \,,
\end{align}
where $m_K$ is the kaon mass and $\Theta(x)$ is the unit step function.
Here, we use the unit step function as a general phase-space correction factor to avoid rotating charged states into more massive counterparts that we do not have phase space for (\eg~rotating pions into kaons below the $KK$ threshold). 
For exclusive states more specific correction factors can be used.
We also obtain an unknown phase factor in \cref{eq:hadronic master} that we take to be $\cos(\phi) = -0.66$ for recasting, since this agrees best with the perturbative limit, see \cref{app:spectral}.

Since our method is based on isospin partners of charged currents,  we do not have access to final states that are isosinglets. 
Here, we argue that their contribution is sub-leading. 
First, in the absence of $G$-parity breaking, states that are eigenstates of $G$-parity can be reached only by either a vector current or an axial current, but not both.  
Since we are assuming isospin symmetry, for an axial-vector $X$ boson final states with an odd number of pions can be reached only through isovector decays, and states with an even number of pions only through isoscalar decays.
This raises the question of whether our leading-order hadronic decay is the isoscalar decay to two pions, or the isovector decay to three pions. 
The decay to two pions would have more phase space, and should dominate over the three-pion final state. 
However, the process $X\to 2\pi$ violates parity, hence only the isovector decay to the $3\pi$ final state is allowed.
%

An additional consideration is the isoscalar component of other decays, such as $X\to[\eta\pi\pi]_{I=0}$. 
We estimate this hadronic rate by considering the corresponding decay of the $f_1$ isoscalar meson, $f_1 \to \eta\pi\pi$, since the $f_1$ has the same \jpc numbers as a purely axial $X$ boson.
This hadronic decay rate for the $f_1$ is obtained~\cite{Rudenko:2017bel} by studying the process 
\begin{align}
    e^+e^-\to f_1\to a_0\pi^0\to\eta\pi^0\pi^0\,.
\end{align} 
If required, the $a_0$ mediated contribution of the charged-pion final state can be obtained assuming isospin symmetry. 
Replacing the $f_1$ with an $X$ boson of the same mass to obtain $\Gamma(X \to a_0\pi^0)$, and using our estimate for the total $X$ hadronic width from \cref{eq:hadronic master}, we estimate $\BR(X \to a_0 \pi^0) \approx 2\%$. 
Therefore, we conclude that the isoscalar component provides an $\cO(\%)$ contribution to the total hadronic rate, which justifies ignoring it in \cref{eq:hadronic master}.
This is similar to the vector current case, where the isoscalar contribution is much smaller than the isovector one, see \eg~\cite{Ilten:2018crw}. 

\section{Experiments} 
\label{sec:experiments}
\subsection{APEX and A1}

The A1~\cite{Merkel:2014avp} and APEX~\cite{Abrahamyan:2011gv} experiments provide electron bremsstrahlung constraints on promptly decaying dark photons. 
The decay $\aprime\to e^+e^-$ was searched for by both experiments in the regime of $m_A\lesssim 300 \,\mev$, which is below the hadronic threshold. 
A1 and APEX searched for a dark photon produced in electron-nucleus fixed-target scattering which then decays promptly to an $e^+e^-$ pair. 
Only promptly decaying dark photons are considered so the efficiency of detection is the same for an $X$ boson if its lifetime is short enough for its decays to be classified as prompt.  
This is not the case for all models; therefore, we must take into consideration lifetime dependencies.  
Recasting is done using
\begin{align}
   g_X^2 \Big[(x^e_V)^2 + (x^e_A)^2\Big] \frac{\BR\left(X\to e^+e^-\right) }{\BR\left(\aprime \to e^+e^-\right) } = \varepsilon^2 e^2 \left(1-e^{\tilde{t}/\tau_X}\right)\,,
\end{align}
where $\tilde{t}$ is the longest proper decay time the $X$ can have and still qualify as prompt~\cite{Ilten:2018crw}.  

\subsection{BaBar}

BaBar searched for a dark photon in the mass region $20\mev \lesssim m_{A'} \lesssim 10\gev$ produced by $e^+e^-$ annihilation and subsequently decaying to an electron-positron or muon-antimuon pair. 
The BaBar collaboration published strong constraints on both visible~\cite{Lees:2014xha} and invisible~\cite{Lees:2017lec} \aprime decays. 
We use \cref{eq:hadronic master} to obtain the hadronic rate for the axial current and we use the framework of \darkcast~\cite{Ilten:2018crw} to obtain the rate for the vector current. 
Altogether, recasting is done using 
\begin{align}
   g_X^2 \Big[(x^e_V)^2 + (x^e_A)^2\Big] \frac{\BR\left(X\to e^+e^-, \mu^+\mu^-\right) }{\BR\left(\aprime \to e^+e^-, \mu^+\mu^-\right) }= \varepsilon^2 e^2 \left(1-e^{\tilde{t}/\tau_X}\right)\,.
\end{align}
%

\subsection{NA64}

The NA64 experiment~\cite{Banerjee:2019pds} set bounds in the \mev--\gev mass region on an invisibly decaying dark photon via the detection of missing energy carried away by hard bremsstrahlung produced in the reaction $eZ\to eZA^\prime$. 
This bremsstrahlung is due to high-energy electrons scattering in a fixed beam-dump target.
Both the ratio of branching fractions and efficiencies are taken to be unity for invisible decays, thus these bounds are easily recast using 
\begin{align}
    g_X^2 \Big[(x^e_V)^2 + (x^e_A)^2\Big] = \epsilon^2 e^2\,.
\end{align}
In addition, NA64$_\mu$~\cite{Gninenko:2019qiv,Sieber:2021fue} is a planned fixed-target experiment in which a dark photon can be produced via muon bremsstrahlung and subsequently decays invisibly. 
Future bounds that will be set by NA64$_\mu$ can be recast by adapting \cref{eq:Abrem} to muon bremsstrahlung and equating
\begin{align}
    \sigma(\mu Z\to \mu Z X) = \sigma(\mu Z\to \mu Z \aprime) \, .
\end{align}
A similar relation holds also for recasting the future results of the M$^3$ experiment~\cite{Kahn:2018cqs}, and from ATLAS as a muon on fixed-target experiment~\cite{Galon:2019owl}. 

\subsection{Beam Dumps}

Limits on dark photons have been set~\cite{Bjorken:2009mm,Andreas:2012mt} using data from the E137, E141, E774, KEK, and Orsay electron beam-dump experiments~\cite{Riordan:1987aw,Bjorken:1988as,Bross:1989mp,Konaka:1986cb,Davier:1989wz}, which were sensitive to decays into electrons and photons. 
Furthermore, limits on the $\aprime\to e^+e^-$ decay were set using data the $\nu$-CAL I~\cite{Blumlein:1990ay,Blumlein:1991xh} proton beam-dump experiment.  
All beam-dump experiments only probe long-lived dark photons. 

The production mechanism for the $X$ particle in these experiments is electron or proton bremsstrahlung. 
The electron beam-dump experiments (E137, E141, E774, KEK, Orsay) set bounds in the regime of $m_A\lesssim 300 \,\mev$, which is below the axial-vector hadronic threshold. 
The proton beam-dump experiment $\nu$-CAL explored slightly beyond the hadronic threshold, setting bounds for $m_A\lesssim 400 \,\mev$. 
We again obtain the efficiency ratios as described in  \cite{Ilten:2018crw}. 
Therefore, recasting the electron beam-dump results requires solving
\begin{align}
    g_X^2 \Big[(x^e_V)^2 + (x^e_A)^2\Big] \mathcal{B}(X \to e^+e^-) \epsilon[\tau_X(g_X)]
	\geq 	(\varepsilon_{\rm max} e)^2 \mathcal{B}(\aprime \to e^+e^-) \epsilon[\tau_{\aprime}(\varepsilon_{\rm max})]  \, ,
\end{align}
while recasting the proton beam-dump constraints requires solving
\begin{align}
 g_X^2 \left[(2x^u_V + x^d_V)^2 + (2x^u_A + x^d_A)^2 \left(\frac{F^A(m_X)}{F^V(m_X)}\right)^2 \right] \mathcal{B}(X \to e^+e^-)  \epsilon[\tau_X(g_X)] \nonumber \\
 \geq (\varepsilon_{\rm max} e)^2 \mathcal{B}(\aprime \to e^+e^-) \epsilon[\tau_{\aprime}(\varepsilon_{\rm max})]  \, .
\end{align}
%

\subsection{LHCb}

The LHCb experiment performed searches for a dark photon produced in proton-proton collisions at a center of mass energy of $13 \,\tev$, and decaying into via $\aprime\to\mu^+\mu^-$ \cite{Aaij:2017rft,Aaij:2019bvg}. 
Both limits on prompt and long-lived \aprime decays were published. 
The prompt \aprime search covers the mass range from the $\mu^+\mu^-$ threshold up to $70 \,\gev$. 
The long-lived search is restricted to the mass region $240\leq m_{\aprime} \leq 350 \mev$. 
Among the different production mechanisms for the dark photon, only Drell-Yan production is relevant for a massive boson with only axial couplings.
The DY production cross section is given in \cref{eq:DY,eq:DYratio}. 
For prompt searches, recasting is done using
\begin{align}
    \sum_{q_i} \frac{\sigma(q_i\bar{q_i}\to \gamma^\ast)} {\sigma(\text{DY}\to\gamma^\ast)} \frac{g_X^2 \Big[(x^q_{V})^2 + (x^q_A)^2\Big]}{e^2\varepsilon^2 Q^2_{q} }
=   \frac{\BR\left(X\to \mu^+\mu^-\right) } 
    {\BR\left(\aprime \to \mu^+\mu^-\right)} \left(1-e^{\tilde{t}/\tau_X}\right)\, .
\end{align}
The vector-meson-mixing-based production mechanisms for the \aprime are excluded for a massive axial vector, as explained in \cref{sec:prod}. 
Therefore, these mechanisms are relevant only for vector currents, so either purely vector bosons or chiral bosons. 
In these cases, we use the mechanism provided in \darkcast \cite{Ilten:2018crw} for 
$\rho,\omega,\phi \to \aprime $ and $\eta \to \aprime \gamma\,,\;\;\omega \to \aprime \pi^0$ to recast the LHCb results.

\subsection{Neutrino Experiments}

We also recast bounds set by CHARM~II~\cite{CHARM-II:1994dzw}, BOREXINO~\cite{Bellini:2011rx} and TEXONO~\cite{TEXONO:2009knm, TEXONO:2006xds, Chen:2014dsa} on the minimal $B-L$ extension of the SM from Ref.~\cite{Bauer:2018onh} (see also relevant discussion in~\cite{Greljo:2022dwn}). 
For this we approximate $\sigma_X \mathcal{A}_X/(\sigma_{BL} \mathcal{A}_{BL}) \approx \sigma_{X}/\sigma_{BL} $, where $\mathcal{A}_X$ is the acceptance. 
The effect of this approximation on the resulting bounds is less than $\cO(1)$.
The bounds are set by measuring the recoil energy of the electron in the elastic-scattering processes  $e^-\nu_\mu \to e^-\nu_\mu$, $e^-\nu_e \to e^-\nu_e$, and $e^-\bar{\nu}_e \to e^-\bar{\nu}_e$ at CHARM~II, BOREXINO, and TEXONO, respectively. 

\section{Example Models}
\label{sec:examples}

\begin{figure}[t]
    \centering
    \includegraphics[width=0.48\textwidth]{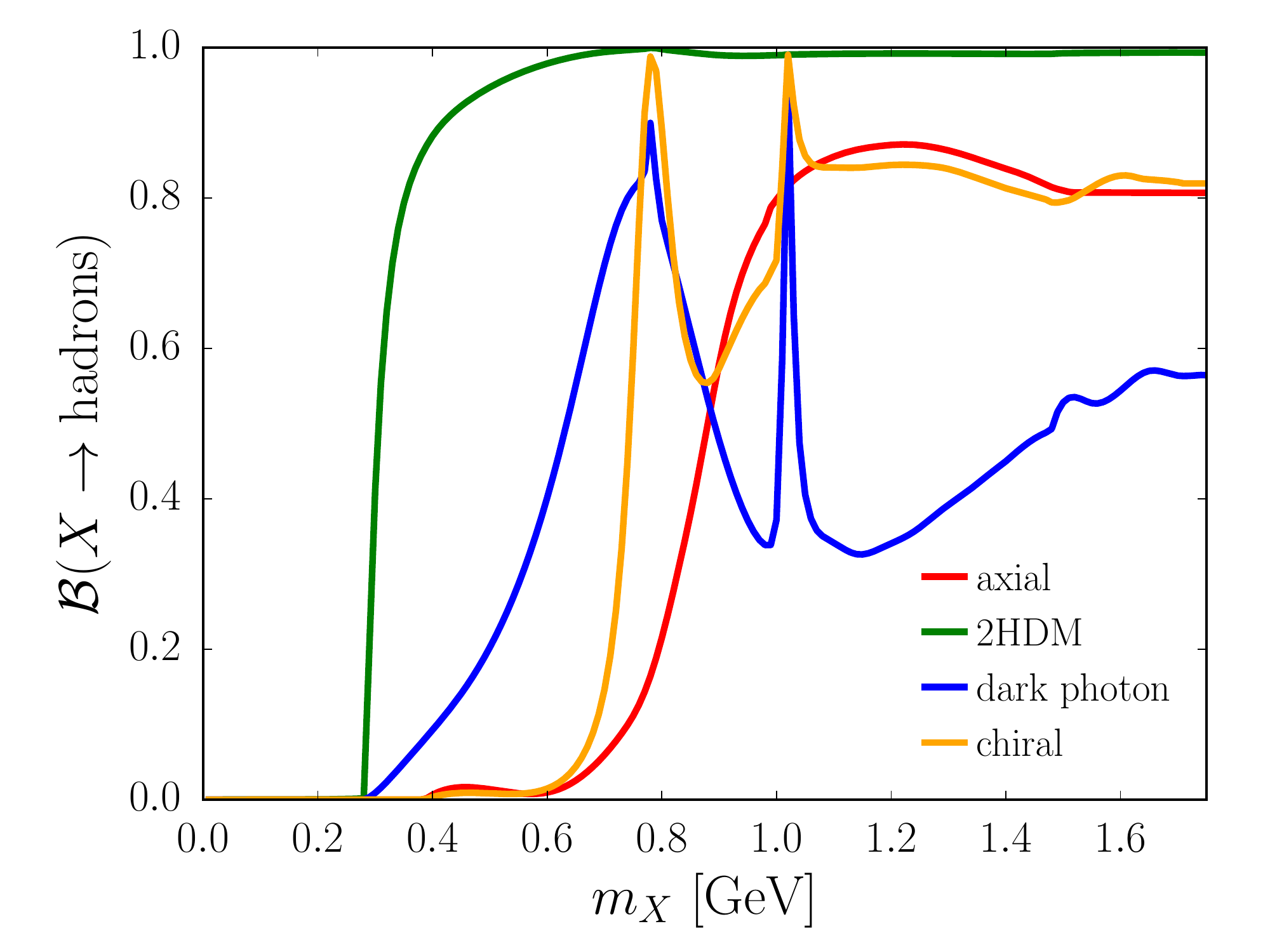}
    \includegraphics[width=0.48\textwidth]{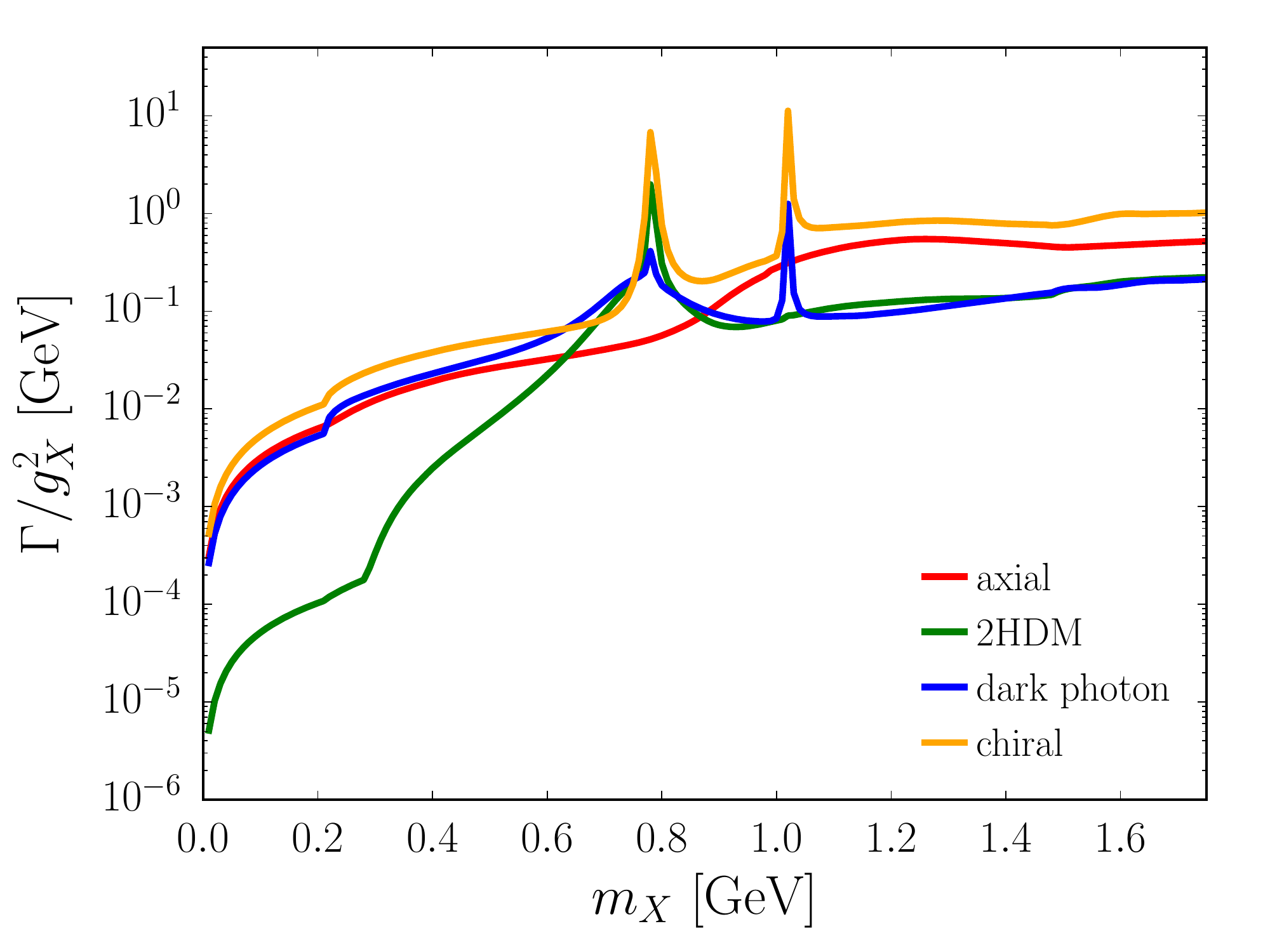}
    \caption{The hadronic branching fraction and total decay width for the dark photon and example models: axial boson, chiral boson and 2HDM. The charges are given in \cref{tab:model charges}. The decay width is normalized by the coupling $g_X$.}
    \label{fig:hadronic bfrac}
\end{figure}

We use the framework developed above, along with the previous work on purely vectorial couplings from~\cite{Ilten:2018crw}, to recast several example models: 
(i)~a purely axial boson model; 
(ii)~a chiral model with both vector and axial couplings; 
and 
(iii)~the two-Higgs-doublet~(2HDM) model from Ref.~\cite{Kahn:2016vjr}, see details below. 
The relevant charges are outlined in \cref{tab:model charges}, 
where we take flavor universal couplings for all three models. 
In all of the models, the branching fraction to dark matter is first taken to be zero, {\em i.e.}\  $\BR\left(X\to\bar{\chi}\chi\right) = 0$, then subsequently the case where decays to dark matter dominate is considered. 
For the latter, the limits for purely invisible decays are independent of the dark matter mass assuming $\frac{m_e}{m_X}$ is small. 
For the visible scenario, the hadronic branching fractions and decay widths for each model, which are obtained following \cref{sec:decay}, are shown in \cref{fig:hadronic bfrac}. 

\begin{table}[t]
    \centering
    \begin{tabular}{|c?c|c|c|c?c|c|c|c|}
        \hline
        & $x_e^V$ & $x_\nu^V$ & $x_{u,c,t}^V$ & $x_{d,s,b}^V$ & $x_e^A$ & $x_\nu^A$ & $x_{u,c,t}^A$ & $x_{d,s,b}^A$ \\
        \hline
        Axial & 0 & 1/4  & 0 & 0 & -1 & -1/4  & 1 & -1 \\
        Chiral & -1 & 0 & 1 & 1 & -1 & 0 & 1 & -1 \\
        2HDM & 0.044 & 0.05  & 1.021 & 0.015 & -0.1 & 0.05  & -0.95 & -0.1 \\
        \hline
    \end{tabular}
    \caption{Charges of the SM fermions under $X$ boson interactions for the example models considered. For simplicity, these models all have flavor universal couplings.
    }
    \label{tab:model charges}
\end{table}
%

\subsection{Current Bounds}
\label{sec:currentbounds}

\begin{figure}[t]
    \centering
    \includegraphics[width=0.91\textwidth]{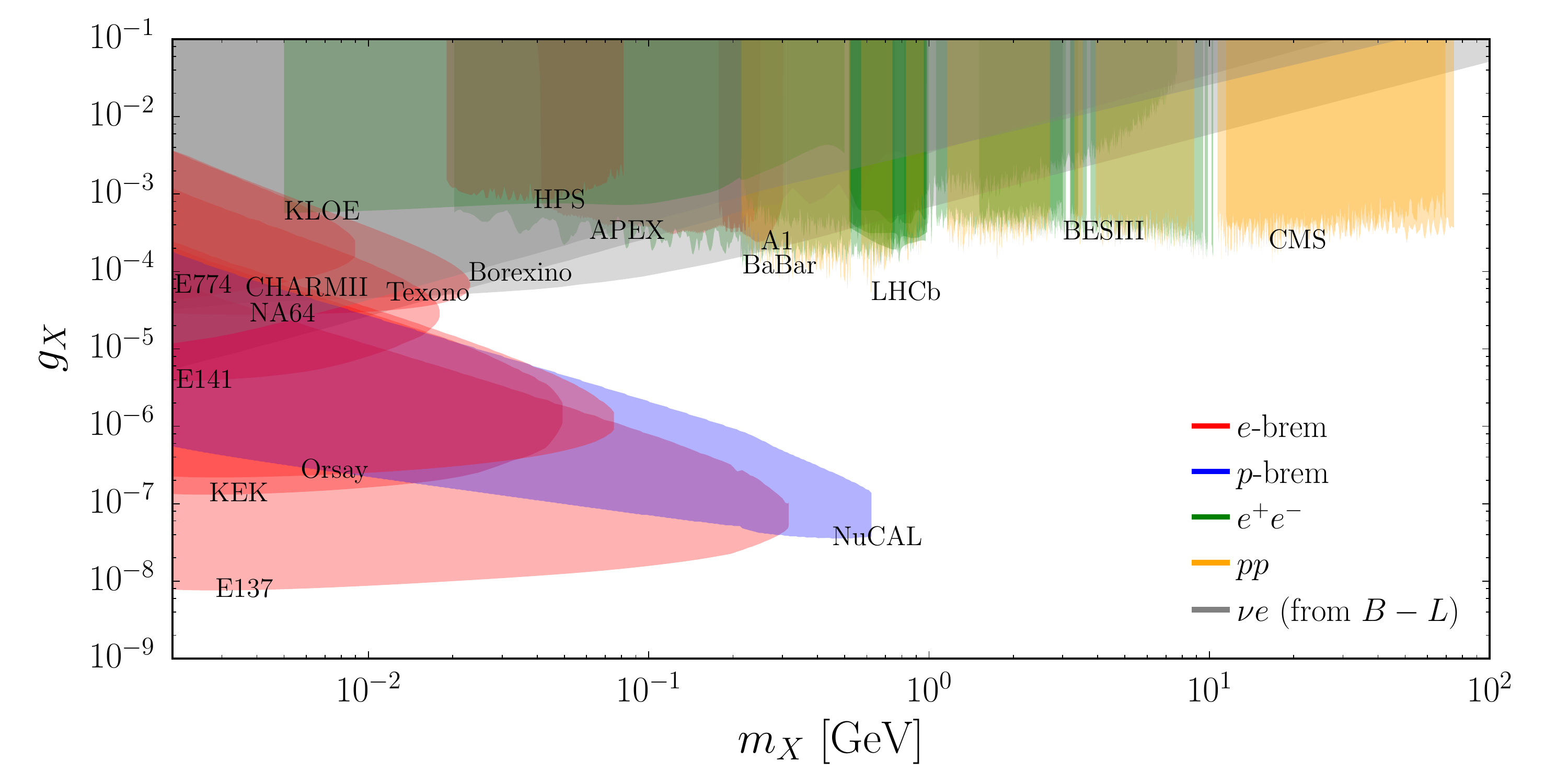}
    \caption{The dark photon bounds recast to a model with a massive axial boson.}
    \label{fig:axialrecast}
\end{figure}

\begin{figure}[t]
    \centering
    \includegraphics[width=0.91\textwidth]{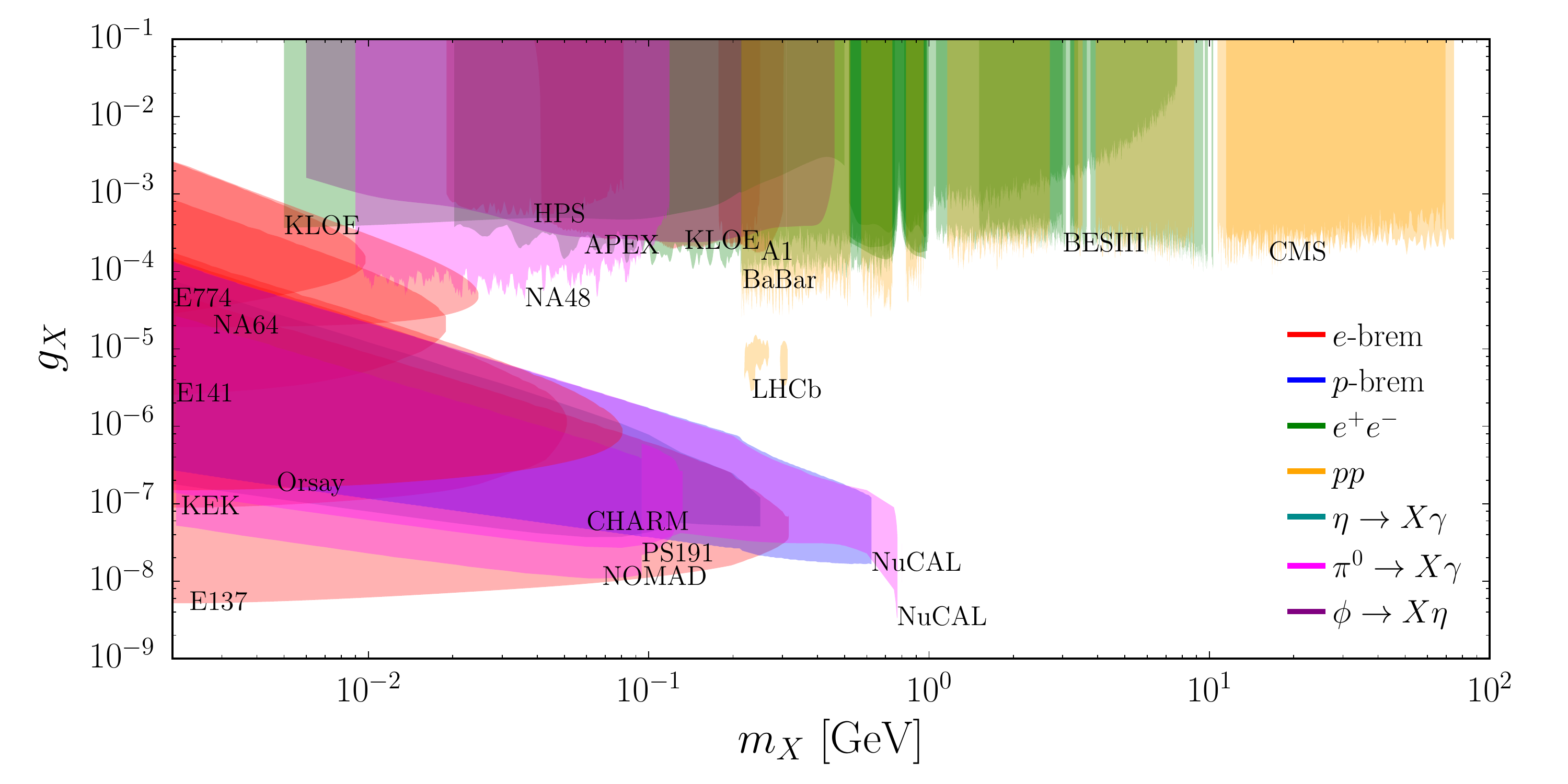}
    \caption{The recast dark photon bounds for a model with a massive boson with both vector and axial couplings to the SM fermions.}
    \label{fig:chiralrecast}
\end{figure}

\begin{figure}[t]
    \centering
    \includegraphics[width=0.91\textwidth]{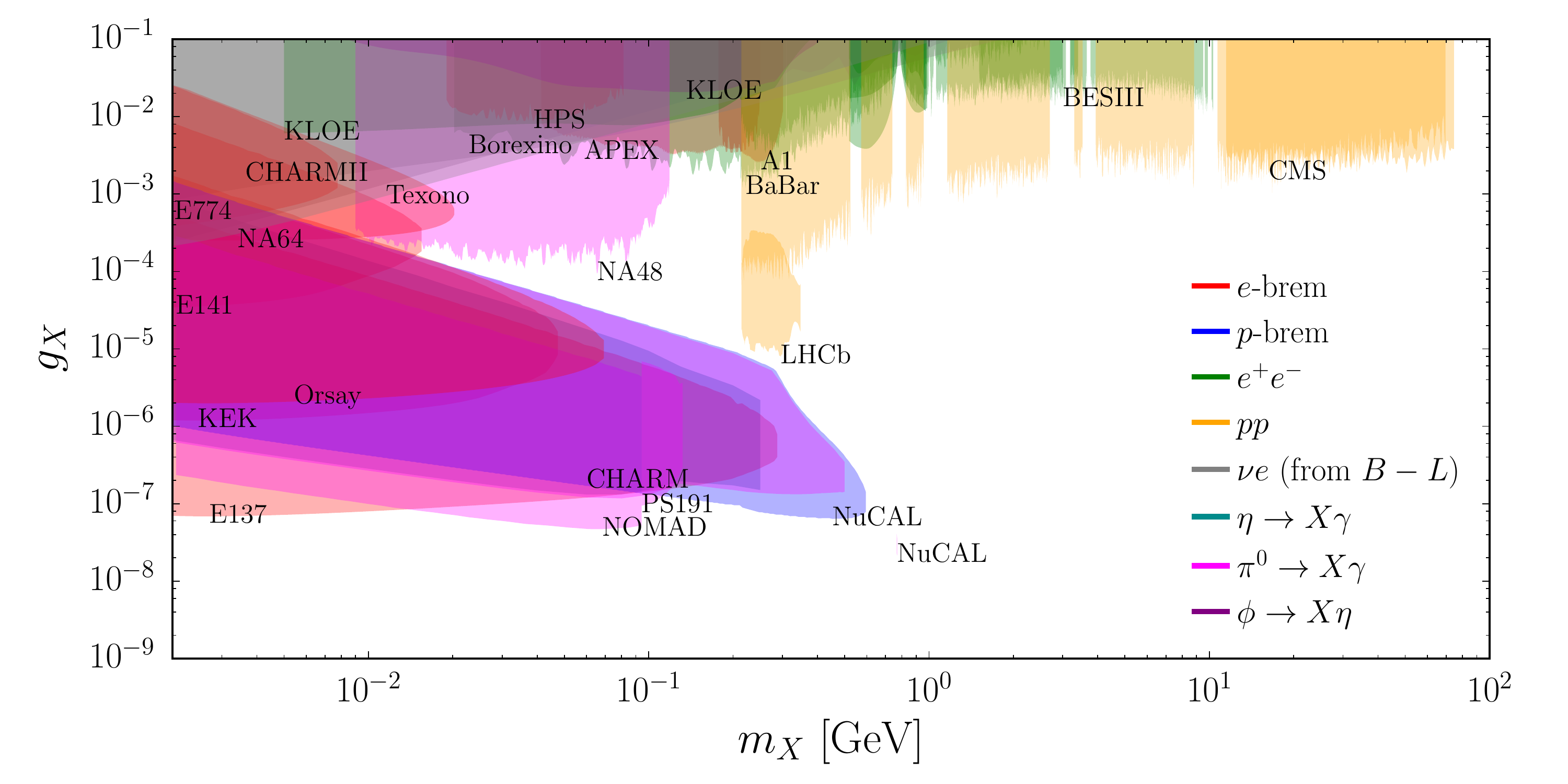}
    \caption{Recast dark photon bounds for a 2-higgs-doublet model with $q_{H_u}=2, q_{H_d}=0.1, \tilde{\theta}_D=0.1$. The coupling for $X$ to the SM fermions as a function of $q_{H_u},q_{H_d}$ and $\tilde{\theta}_D$ can be seen in \cref{eq:2hdmchargeV,eq:2hdmchargeA}.}
    \label{fig:2hdmrecast}
\end{figure}

\begin{figure}[t]
    \centering
    \includegraphics[width=0.91\textwidth]{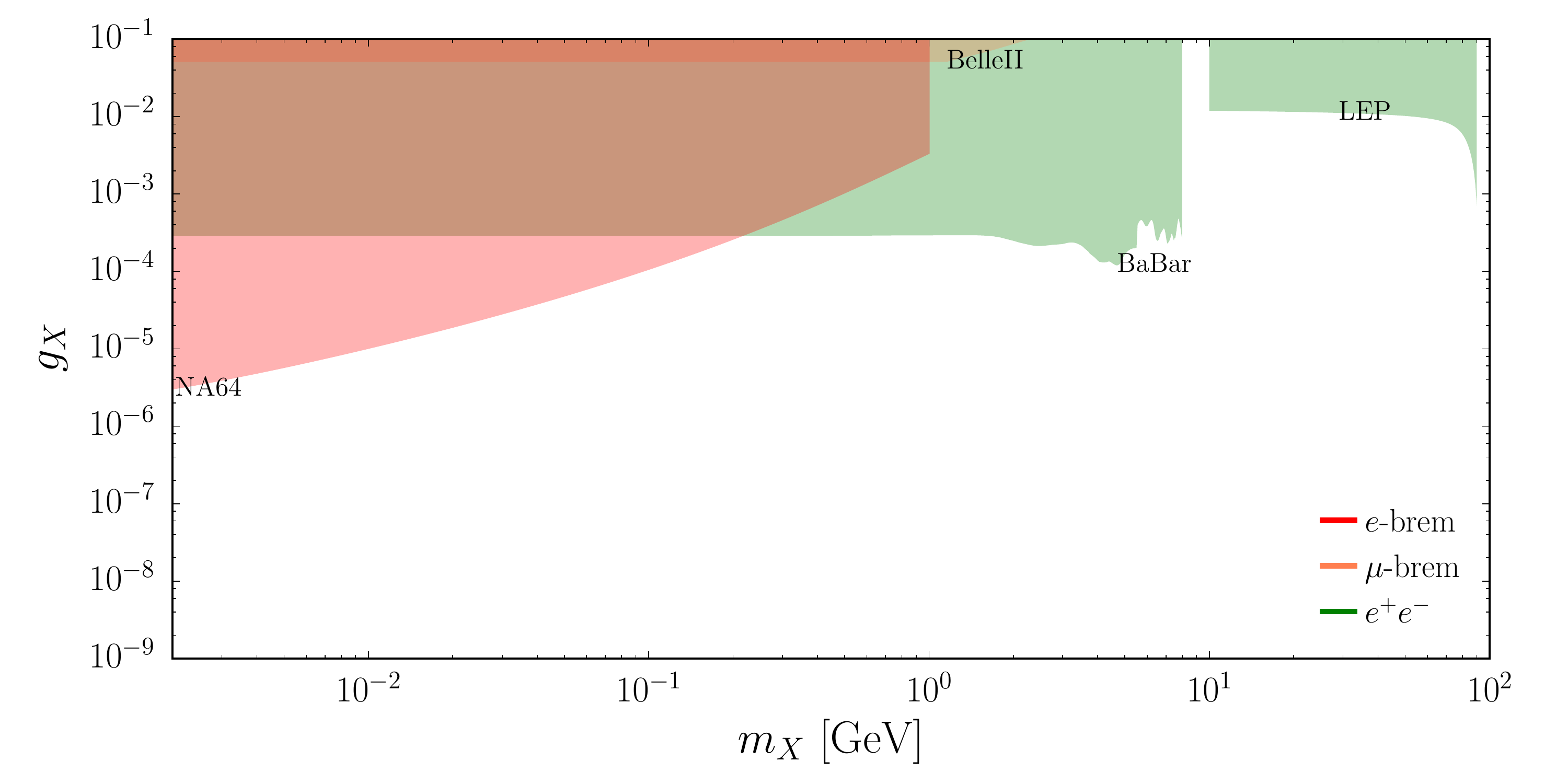}
    \caption{The invisible dark photon bounds recast to a model with a massive axial boson.}
    \label{fig:invisibleaxialrecast}
\end{figure}

\begin{figure}[t]
    \centering
    \includegraphics[width=0.91\textwidth]{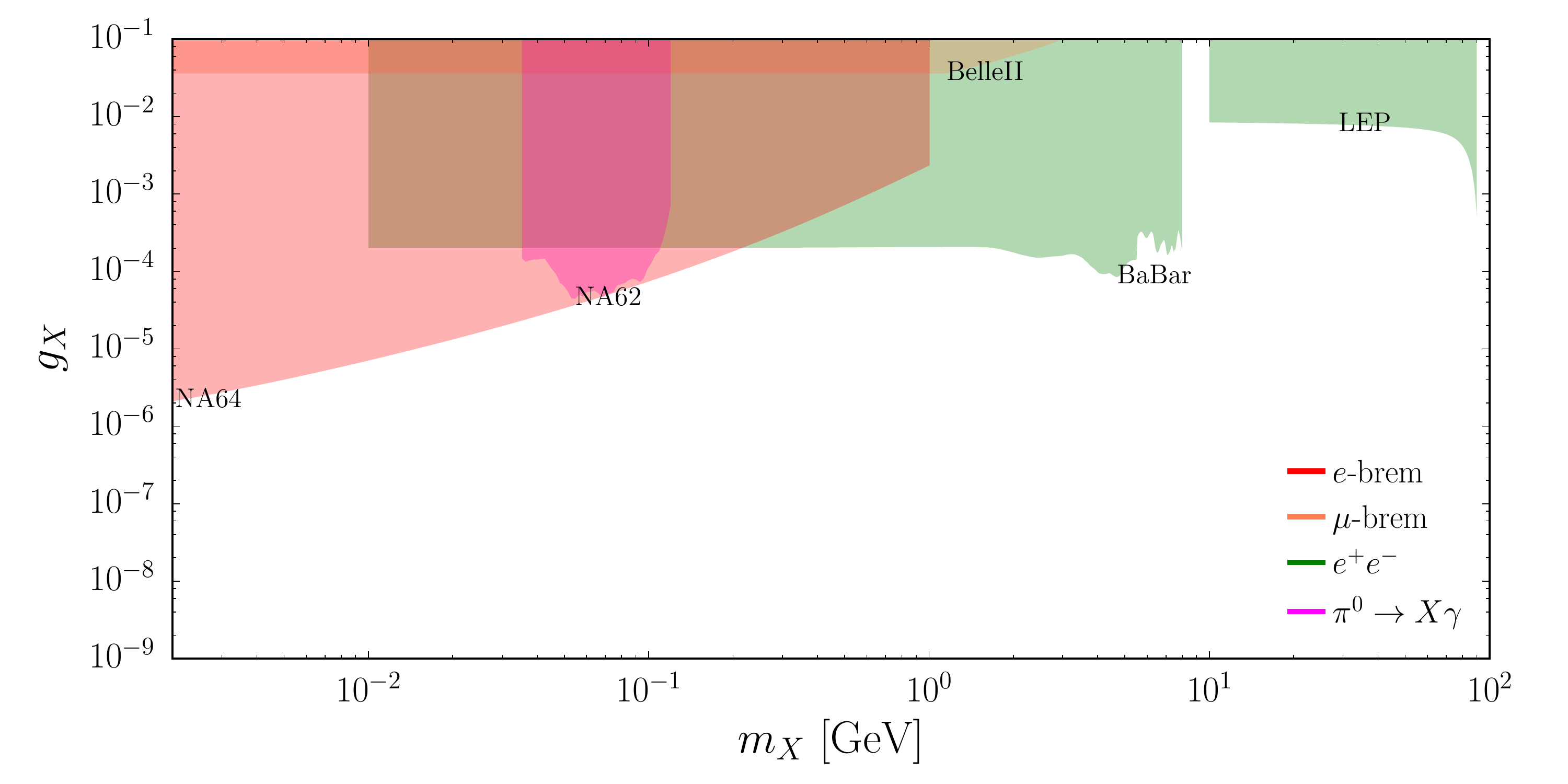}
    \caption{The recast invisible dark photon bounds for a model with a massive boson with both vector and axial couplings to the SM fermions.}
    \label{fig:invisiblechiralrecast}
\end{figure}

\begin{figure}[t]
    \centering
    \includegraphics[width=0.91\textwidth]{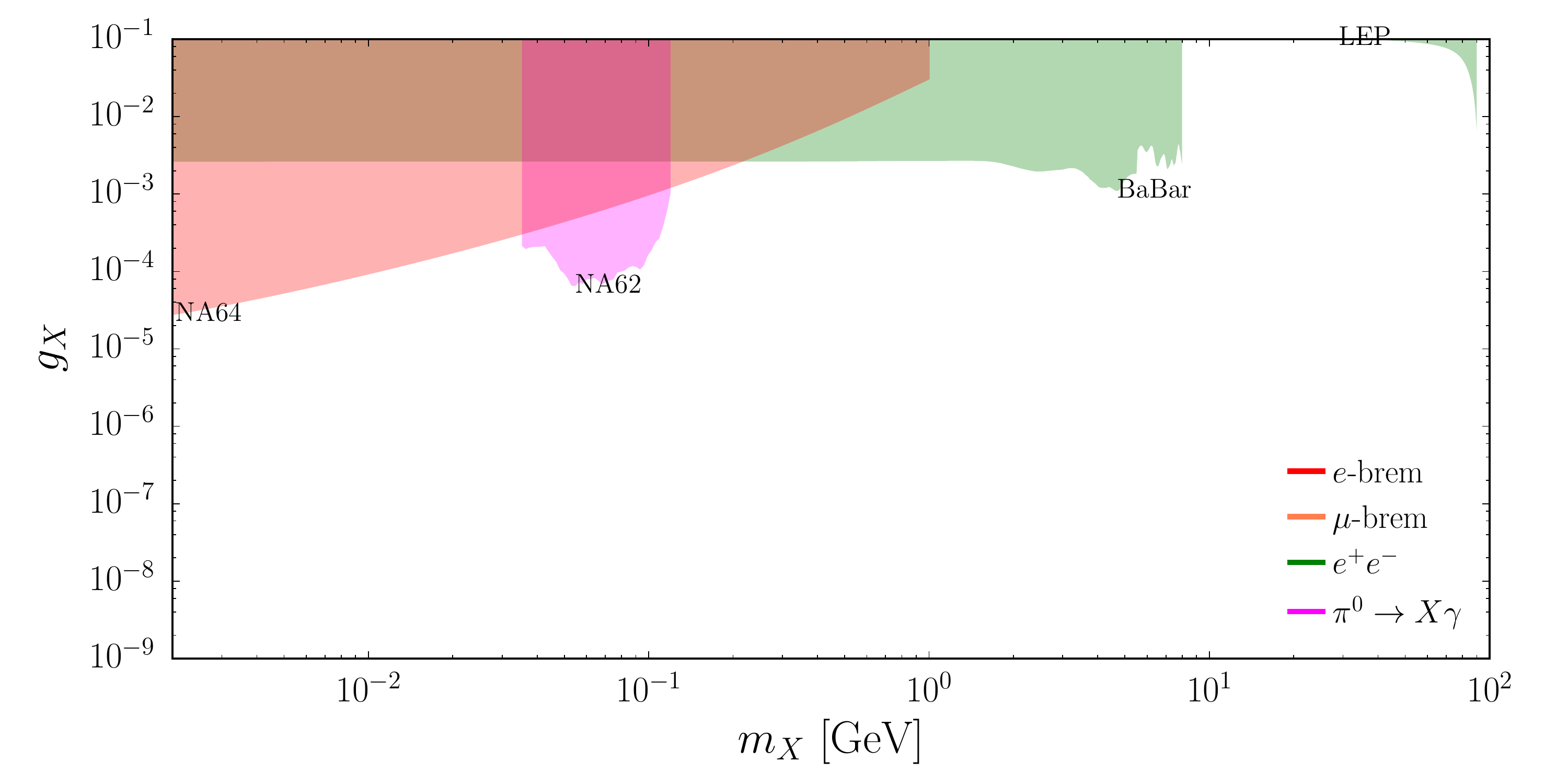}
    \caption{Recast invisible dark photon bounds for a 2-higgs-doublet model with $q_{H_u}=2, q_{H_d}=0.1, \tilde{\theta}_D=0.1$. The coupling for $X$ to the SM fermions as a function of $q_{H_u},q_{H_d}$ and $\tilde{\theta}_D$ can be seen in \cref{eq:2hdmchargeV,eq:2hdmchargeA}.}
    \label{fig:invisible2hdmrecast}
\end{figure}

The recast dark-photon bounds, obtained following \cref{sec:experiments}, can be seen in \cref{fig:axialrecast} for the axial model, in \cref{fig:chiralrecast} for the  chiral model, and in \cref{fig:2hdmrecast} for the 2HDM for the visible-decay scenario.
The invisible-decay bounds, {\em i.e.}\ assuming decays to dark matter are kinematically allowed and dominant, are shown in \crefrange{fig:invisibleaxialrecast}{fig:invisible2hdmrecast}.

\subsection{Projections}
\label{sec:projections}

\begin{figure}[!h]
    \centering
    \includegraphics[width=0.91\textwidth]{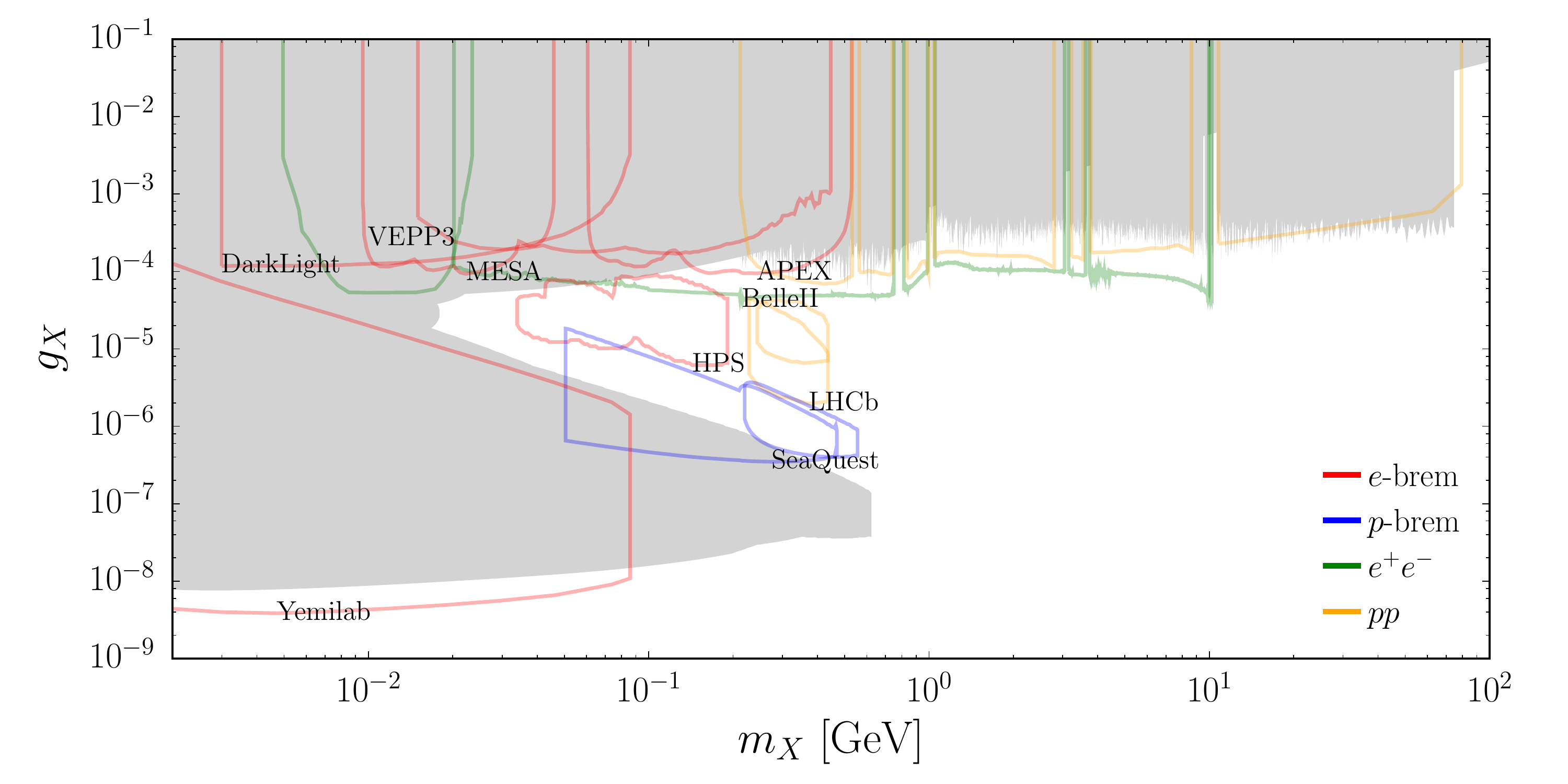}\\
    \includegraphics[width=0.91\textwidth]{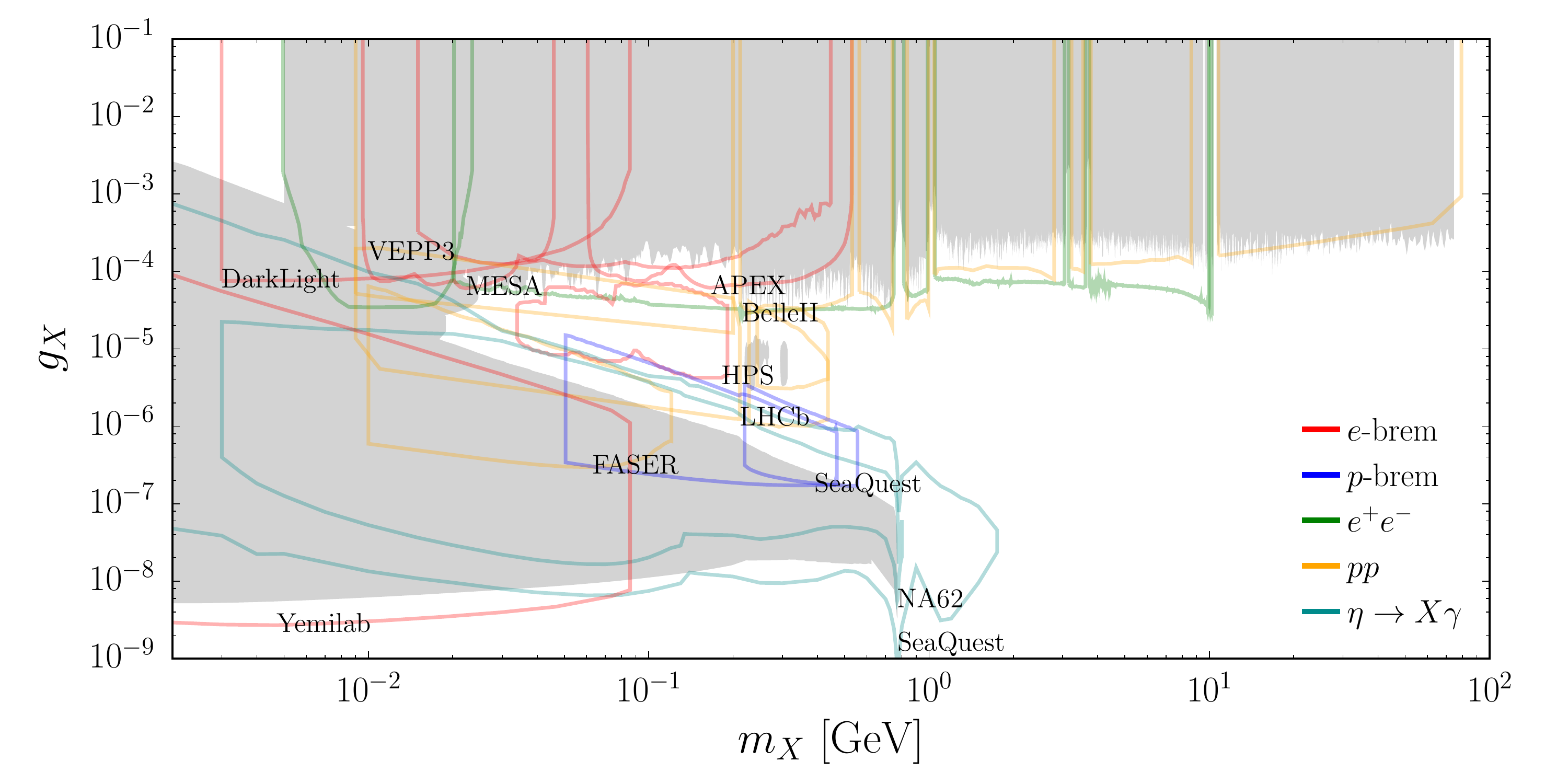}\\
    \includegraphics[width=0.91\textwidth]{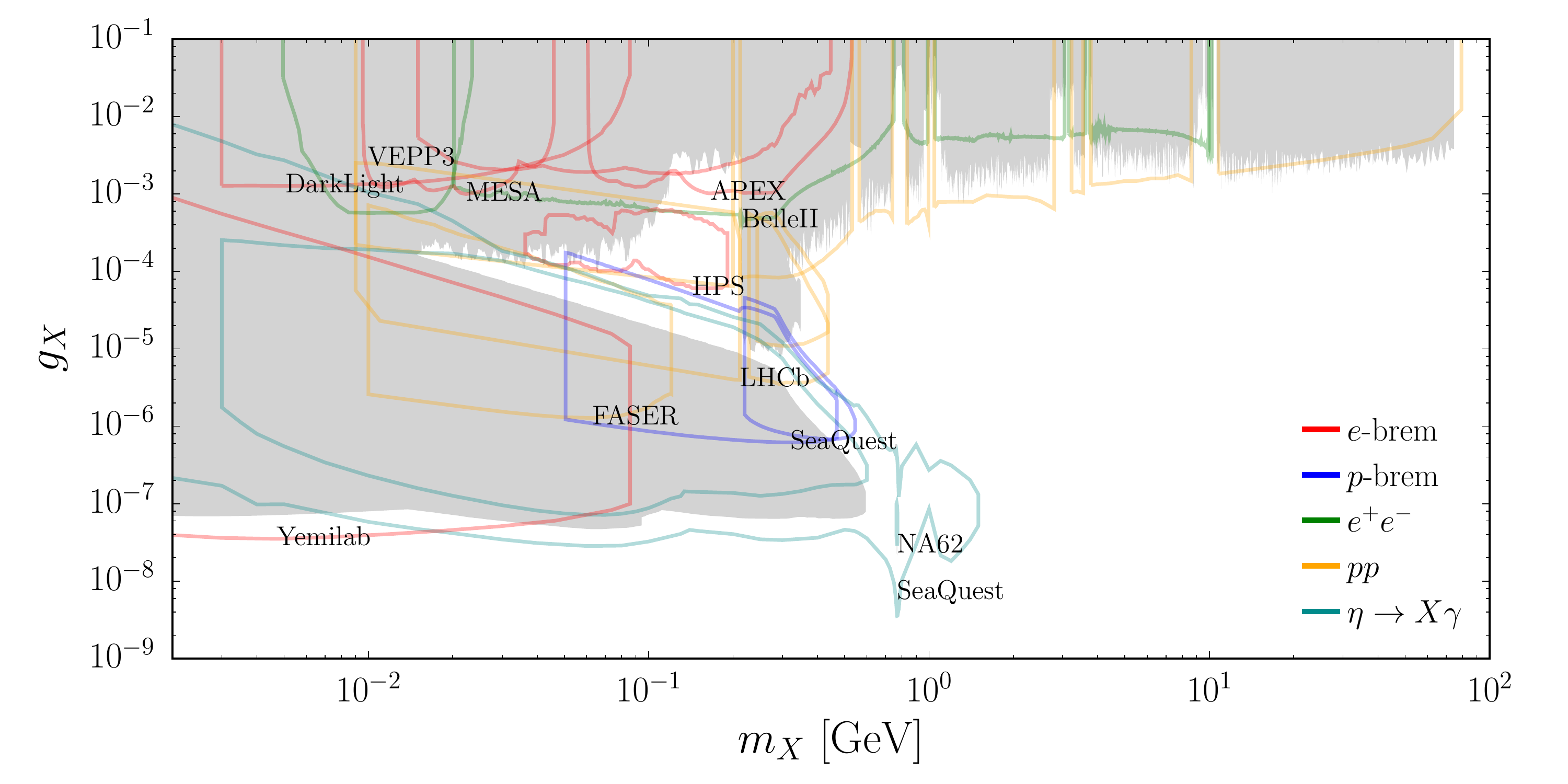}
    \caption{The projected bounds for the example models: axial boson, chiral boson, and 2HDM.}
    \label{fig:projections}
\end{figure}

We now derive future sensitivities for the axial model, chiral model, and the 2HDM using the following projections for dark photon searches: APEX~\cite{Essig:2010xa}, Belle II~\cite{Belle-II:2018jsg}, DarkLight~\cite{Kahn:2012br}, FASER~\cite{FASER:2018eoc}, HPS~\cite{HPS:2016jta}, LHCb~\cite{Ilten:2015hya,Ilten:2016tkc,Craik:2022riw}, MESA~\cite{Beranek:2013yqa}, NA62~\cite{Tsai:2019mtm}, SeaQuest~\cite{Gardner:2015wea}, VEPP-3~\cite{Wojtsekhowski:2012zq}, and Yemilab~\cite{Seo:2020dtx}. 
Projections from Mu3e~\cite{Echenard:2014lma} are not included as the $\mu$-bremsstralhung approximation is not expected to hold within the relevant mass range. 
The recast FASER, LHCb, NA62, and SeaQuest projections using the \aprime production mechanisms of $\pi^0\to\aprime\gamma$ or $\eta\to\aprime\gamma$ do not contribute to models with only axial couplings, \textit{i.e.}~the axial model considered here, but do contribute to models that also have vector couplings. 
The production and decay mechanisms are summarized in \cref{tab:projectedEx} and the projected bounds for each of the example models outlined in \cref{sec:currentbounds} are shown in \cref{fig:projections}.

\begin{table}[t]
    \centering
    \begin{tabular}{|c|c|c|}
        \hline
         \,& production & decay \\
         \hline
         APEX & $e$-bremsstrahlung & $e^+e^-$ \\
         Belle II & $e^+e^-$ & $e^+e^-, \mu^+\mu^-$ \\
         DarkLight & $e-$bremsstrahlung & $e^+e^-$ \\
         FASER & meson decays & $e^+e^-$ \\
         HPS & $e-$bremsstrahlung & $e^+e^-$ \\
         LHCb & DY, meson decays & $e^+e^-, \mu^+\mu^-$ \\
         MESA & $e-$bremsstrahlung & $e^+e^-, \mu^+\mu^-$ \\
         NA62 & $p-$bremsstrahlung, meson decays & $e^+e^-, \mu^+\mu^-$ \\
         SeaQuest & $p-$bremsstrahlung, meson decays & $e^+e^-, \mu^+\mu^-$ \\
         VEPP3 & $e^+e^-$ & $e^+e^-$ \\
         Yemilab & $e-$bremsstrahlung & $e^+e^-$ \\
         \hline
    \end{tabular}
    \caption{Summary of experiments which provide projected bounds for the \aprime, with their relevant production and decay mechanisms.}
    \label{tab:projectedEx}
\end{table}

\newpage

\section{Summary}
\label{sec:summary}

In this work, we have explored new spin-1 bosons with axial couplings to the SM fermions. 
We developed a data-driven method to estimate their hadronic decay rates based on data from $\tau$ decays and using SU(3)$_{\rm flavor}$ symmetry. 
We derived the current and future constraints from the relevant intensity-frontier experiments on several benchmark models, namely a pure axial vector, a chiral model, and a 2-Higgs-doublet model. 
Our framework is generic and can be used to derive the constraints on models with arbitrary vectorial and axial couplings to quarks.
In addition, the code required to reproduce all of our results has been incorporated into the \darkcast package, see  \url{https://gitlab.com/darkcast/releases}. 
We note that our hadronic rate prediction can be systematically improved by more accurate data of the $\tau$ spectral functions, in particular the strange spectral function. In addition it can be improved by precise measurement of low energy parity violating asymmetries in $e^+e^-$ collisions. 
Finally, the results of this study are important not only to help guide searches for new bosons, but also for indirect searches for dark matter. 
For example, the dark-matter annihilation rate to SM particles via an axial mediator can be estimated using our hadronic-rate calculations, see~\cite{Plehn:2019jeo} for the case of vector mediators.  

\begin{acknowledgments}
We thank Zoltan Ligeti for many useful discussions, especially in the early stages of this project, and for providing constructive comments on the manuscript.
We also thank Iftah Galon and Jure Zupan for providing useful feedback.  
The work of CB, YS and MW is supported by NSF-BFS (grant no. 2018683).
In addition, CB and YS are supported by grants from the ISF (No. 482/20), the BSF (No. 2020300) and by the Azrieli foundation. 
MW is also supported by NSF-PHY-1912836. 
\end{acknowledgments}

\appendix

\section{Hadronic Width}
\label{app:hadronicwidth}

We derive an expression for the hadronic decay width using the $\SUF$ relation between the hadronic $\tau$ decay and $e^+e^-$ scattering.
\begin{align}
    \sigma\left(e^+e^-\to \cF^0\right) \approx \frac{g_X^4 (x_A^e)^2}{4\pi s} a_1^{(s)}(s)_{\cF^-}\,.
\end{align}
To construct a master equation for the hadronic rate of $X$ we use an orthogonal basis on which we can project any type of mediating current. Our three basis elements are $\rho$-like, $\omega$-like, and $\phi$-like:
\begin{equation}
    T_{\rho} = \frac{1}{2}
    \begin{pmatrix}
        1 & 0 & 0 \\ 0 & -1 & 0 \\ 0 & 0 & 0
    \end{pmatrix}\, , \qquad
    T_{\omega} = \frac{1}{2}
    \begin{pmatrix}
        1 & 0 & 0 \\ 0 & 1 & 0 \\ 0 & 0 & 0
    \end{pmatrix}\, , \qquad
    T_{\phi} = \frac{1}{\sqrt{2}}
    \begin{pmatrix}
        0 & 0 & 0 \\ 0 & 0 & 0 \\ 0 & 0 & 1
    \end{pmatrix}\, .
\end{equation}
The ${\rho}$-like can be obtained via the non-strange spectral function and we set the $\omega$-like contribution to zero as motivated in the main text. 

The ${\phi}$-like contribution is  a linear combination of the non-strange and strange spectral functions. 
This is due to the way we rotate the charged and neutral hadronic states using $\SUF$:
\begin{align}
    u\bar{d} \to \frac{1}{2} (u\bar{u} - d\bar{d}) \, , \qquad
    u\bar{s} \to \frac{1}{2} (u\bar{u} - s\bar{s}) \, .
\end{align}
To construct the linear combination for the ${\phi}$-like contribution, we define
\begin{align}
    \left|\frac{1}{2} \mel{0}{u\bar{u}-d\bar{d}}{h}\right|^2 = A_{ud}^2 \,, \quad 
    \left|\frac{1}{2} \mel{0}{u\bar{u}+d\bar{d}}{h}\right|^2 = 0 \, ,\quad 
    \left|\frac{1}{2} \mel{0}{u\bar{u}-s\bar{s}}{h}\right|^2 = A_{us}^2 \, ,
\end{align}
where $A^2_{ud}$ and $A^2_{us}$ are related to the spectral functions $a_1(s)$ and $a_1^s(s)$, respectively, as shown below, and $\ket{h}$ is some hadronic state.
We take the square root and thus introduce an unknown phase factor, $\phi$:
\begin{equation}
    \begin{split}
        \mel{0}{u\bar{u}-d\bar{d}}{h} = 2A_{ud} \, ,\quad
        \mel{0}{u\bar{u}+d\bar{d}}{h} = 0 \, ,\quad
        \mel{0}{u\bar{u}-s\bar{s}}{h} = 2e^{i\phi}A_{us} \, .
    \end{split}
\end{equation}
Therefore, $\mel{0}{u\bar{u}}{h} = -\mel{0}{d\bar{d}}{h} = A_{ud}$ and $\mel{0}{s\bar{s}}{h} = A_{ud} -2e^{i\phi}A_{us}$. 
To express $\mel{0}{s\bar{s}}{h}$ in terms of the spectral functions, we have to square it and integrate over phase space: 
\begin{equation}
\begin{split}
    \int d\Phi \left|\mel{0}{s\bar{s}}{h}\right|^2 
    &= \int d\Phi \left(\left|A_{ud} -2e^{i\phi}A_{us}\right|^2\right) \, .
\end{split}
\end{equation}
The squared terms are the spectral functions:
\begin{align}
        \int d\Phi |A_{ud}|^2 &= \int d\Phi\left|\frac{1}{2} \mel{0}{u\bar{u}-d\bar{d}}{h}\right|^2 = \frac{1}{4} a_1(s) \, ,\\
        \int d\Phi |A_{us}|^2 &= \int d\Phi\left|\frac{1}{2} \mel{0}{u\bar{u}-s\bar{s}}{h}\right|^2 = \frac{1}{4} a_1^s(s) \, .
\end{align}
The interference term  we approximate as follows:
\begin{align}
    \cos(\phi) \int d\Phi  A_{ud}A_{us} 
    \approx 
    \cos(\phi) \sqrt{a_1(s)a_1^s(s)} \, .
\end{align}
Our expression for the decay width to a neutral hadronic state $\cF^0$, as in \cref{eq:gamma_hadrons}, is
\begin{equation}
    \Gamma\left(X\to h^0\right) = \Gamma\left(X\to\mu^+\mu^-\right) \left(R_{X}^{\rho} + R_{X}^{\omega} + R_{X}^{\phi}\right)(m_X)\, ,
\end{equation}
where $R_{X}^{\rho}$, $R_{X}^{\omega}$, and $R_{X}^{\phi}$ are the $\rho$-like, $\omega$-like, and $\phi$-like contributions to $R_{\mu}^{X}$, the $X$ mediated equivalent of $R_\mu$.
We define the charge matrix for $X$
\begin{equation}
    Q_A^X = \begin{pmatrix}
    x_A^u & 0 & 0 \\ 0 & x_A^d & 0 \\ 0 & 0 & x_A^s
    \end{pmatrix}\, .
\end{equation}
Therefore, we can write
\begin{equation}
    \begin{split}
        R^{\rho}_{X} 
        &= \left(\Tr[T_{\rho}Q_A^X]\right)^2 R^{\rho}_{\mu} 
        = (x_A^u-x_A^d)^2 R^{\rho}_{\mu} \, ,\\
        R^{\omega}_{X} 
        &= \left(\Tr[T_{\omega}Q_A^X]\right)^2 R^{\omega}_{\mu} 
        = (x_u+x_d)^2 R^{\omega}_{\mu} \, ,\\
        R^{\phi}_{X} 
        &= \left(\Tr[T_{\phi}Q_A^X]\right)^2 R^{\phi}_{\mu} 
        = (x_A^s)^2 R_{\mu}^{\phi} \, .
    \end{split}
\end{equation}
We use the non-strange spectral function in \cref{eq:tau decay to scattering} to express the $\rho$-like contribution as
\begin{align}
    R_{\mu}^{\rho} = \frac{\sigma\left(e^+e^-\to X\to \cF^0 \right)}{\sigma\left(e^+e^-\to X\to \mu^+\mu^-\right)} =  \frac{g_X^4 (x_A^e)^2}{4\pi s}  \frac{a_1(s)_{\cF^-}}{\sigma\left(e^+e^-\to X\to \mu^+\mu^-\right)} \, .
\end{align}
Using the linear combination constructed above the $\phi$-like contribution to the hadronic width is
\begin{equation}
    R_{\mu}^{\phi}  =  \frac{g_X^4 (x_A^e)^2}{4\pi s}\, 
    \left[
    \frac{\frac{1}{4} a_1(s) + a_1^s(s) - \cos(\phi) \sqrt{a_1(s)a_1^s(s)}}{\sigma\left(e^+e^-\to X\to \mu^+\mu^-\right)} \right] \, .
\end{equation}
Together with the step function to account for phase space, we obtain
\begin{align}
    \Gamma_{X \to \mathrm{hadrons}} 
=  \frac{g_X^2 m_X}{4\pi} \Bigg[&(x^u_A-x^d_A)^2 a_1(m_X^2) 
    + (x^s_A)^2 \,      \Theta(m_X^2-4m_K^2) \nonumber\\
&\times\left(\frac{1}{4} a_1(m_X^2) + a_1^s(m_X^2) - \cos(\phi) \sqrt{a_1(m_X^2)a_1^s(m_X^2)}\right)\Bigg] \,.
\end{align}

\section{Spectral functions for tau decays}
\label{app:spectral}

The spectral functions from the ALEPH LEP collaboration~\cite{Davier:2005xq} are obtained by dividing the normalized invariant mass-squared  distribution $(1/N_{V/A})(\deriv{}{N_{V/A}}/\deriv{}{s})$ for a given hadronic mass 
$\sqrt{s}$ by the appropriate kinematic factor
\begin{align}
   v_1(s)/a_1(s) 
   =& 
   \frac{m_\tau^2}{6|V_{CKM}|^2 S_{EW}}
              \frac{\BR(\tau^-\to {V^-/A^-}\nu_\tau)}{\BR(\tau^-\to e^-\bar{\nu}_e\nu_\tau)}
              \frac{1}{N_{V/A}}
              \frac{\deriv{}{N_{V/A}}}{\deriv{}{s}} \\
              &\times \left[ \left(1-\frac{s}{m_\tau^2}\right)^{2} \left(1+\frac{2s}{m_\tau^2}\right)\right]^{-1}\,, \nonumber
\end{align}
where $S_{EW}$ accounts for electroweak radiative  corrections. 

The measured strange $\tau$ spectral function from the ALEPH LEP collaboration~\cite{Davier:2005xq} is the sum of the strange vector spectral function and the strange axial spectral function.
To isolate the axial part of the spectral function, we use the analysis of \cite{ALEPH:1999uux}. 
All the information obtained on the strange decay fractions concerning their $V$/$A$ character are summarized in \cref{fig:VA_Separation}. 
The branching fractions for the $K4\pi$ and $K5\pi$ modes are obtained from the measured branching ratios for the $5\pi$ and $6\pi$ final states, with the relevant Cabibbo suppression and kinematic factors included. 
Vector and axial vector currents in the $K3\pi$, $K4\pi$ and $K5\pi$  are assumed to contribute equally. 

For both the light and strange spectral functions, the sum of the axial and vector perturbative values is taken as $1.1$, calculated at next-to-leading order in QCD. For the strange spectral functions, the perturbative axial and vector values are determined from the analysis of \cite{ALEPH:1999uux}, as described above. The perturbative value for the light spectral vector function is determined by fitting the perturbative behaviour of the hadronic width calculated using $\mathcal{R}_\mu$ to the hadronic width calculated using the spectral functions. This fit is performed simultaneously for both a vector-only model with just a strange-quark coupling and the dark-photon model. Both the perturbative value of the light spectral function and $\cos(\phi)$ are allowed to float in this fit. The perturbative value for the light axial spectral function is then set as $1.1 - v_1$. Both these values match within the uncertainty of the data of \cref{fig:spectral}.

\begin{figure}[h]
    \centering
    \includegraphics[width=0.48\textwidth]{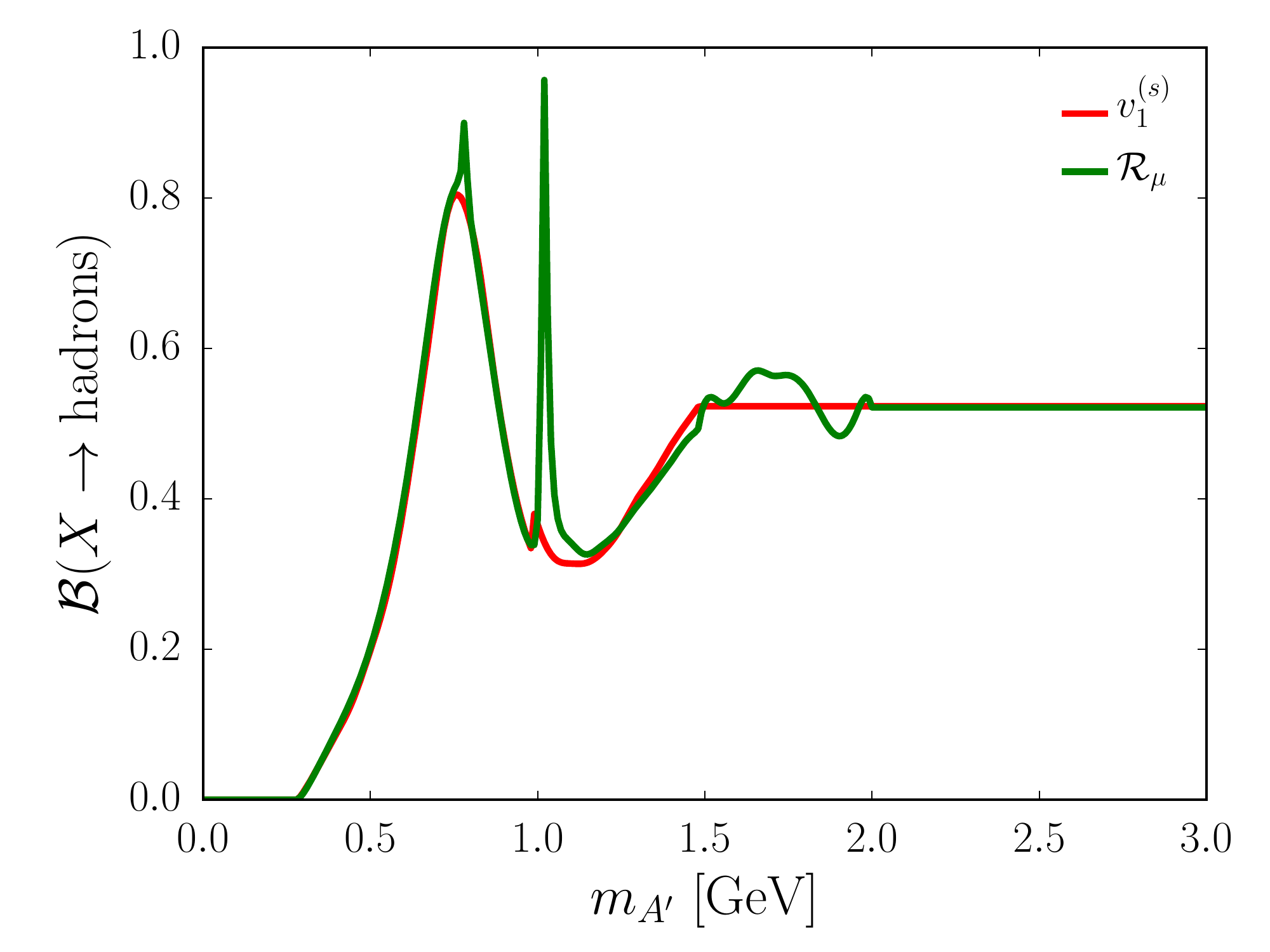}
    \includegraphics[width=0.48\textwidth]{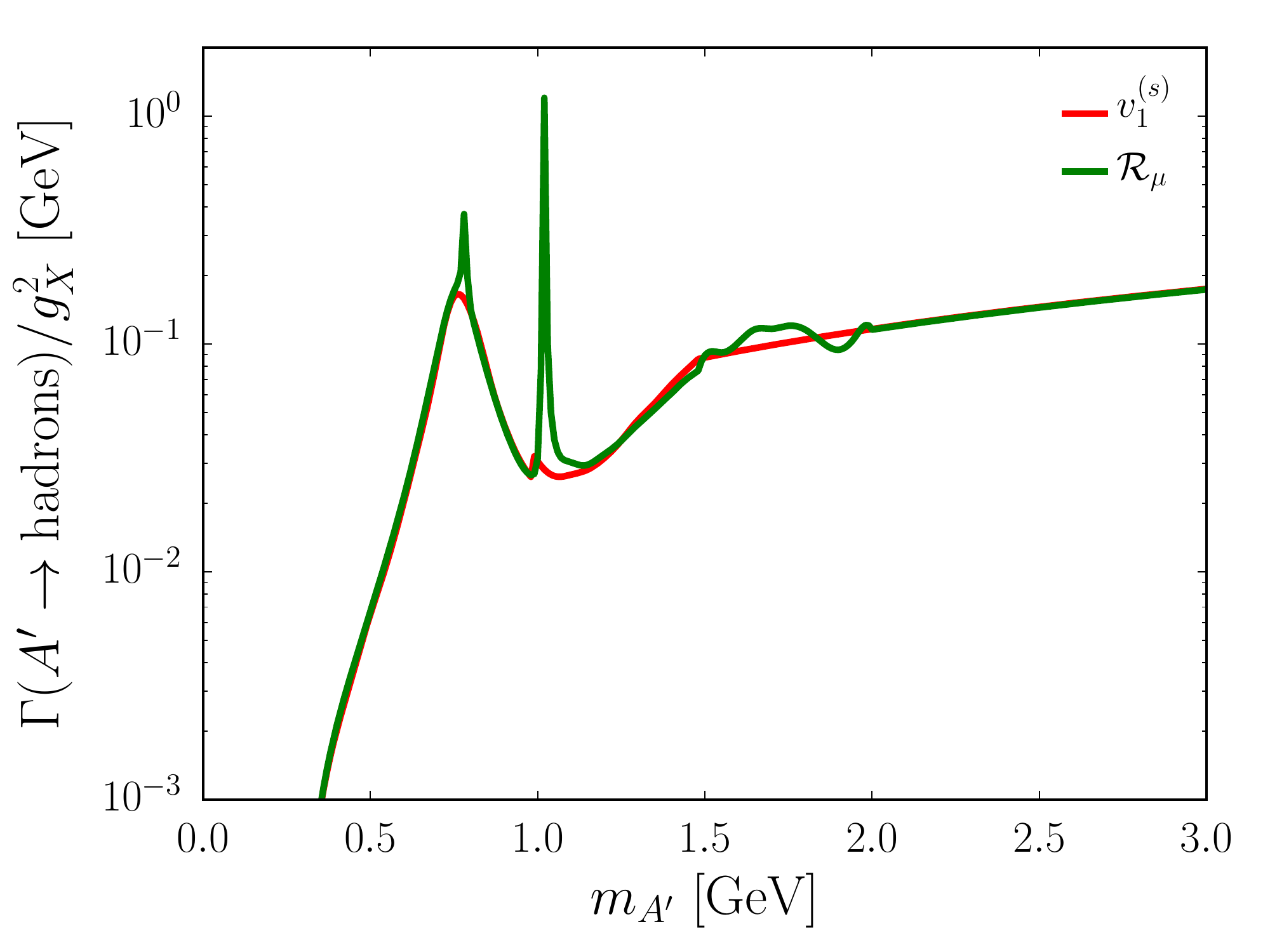}
    \caption{Cross check of the hadronic dark photon branching fraction and width compared between using the vector spectral functions, $v_1$ and $v_1^s$, and the standard \darkcast calculation with $\mathcal{R}_\mu$.}
    \label{fig:check}
\end{figure}

To cross check the extraction of the spectral functions, the hadronic dark photon branching fraction and width calculated using the vector spectral function $v_1$ is compared to the default \darkcast calculation using $\mathcal{R}_\mu$ in \cref{fig:check}. Note that the isoscalar resonances of the $\omega$ and $\phi$ are present in the $\mathcal{R}_\mu$ data but not the vector spectral function. However, the $\rho$ peak and perturbative behaviour is consistent between the two calculations.

\begin{table}[t!]
    \centering
    \begin{tabular}{|c?c|c|c|}
    \hline
        Mode & $B_V (10^{-3})$ & $B_A (10^{-3})$ & $B_{V+A} (10^{-3})$ \\
        \hline
        $K^-$ & - & $6.96\pm 0.29$ & $6.96\pm 0.29$ \\
        $\left(\bar{K}\pi\right)^-$ & $13.60\pm 0.62$ & - & $13.60\pm 0.62$ \\
        $\left(\bar{K}2\pi\right)^-$ & $1.39^{+1.30}_{-1.01}$ & $4.58^{+1.23}_{-1.49}$ & $5.97\pm 0.73$\\
        $K_1^-(1270)\to K^-\omega$ & - & $0.67\pm 0.21$ & $0.67\pm 0.21$ \\
        $K^-\eta$ & $0.29^{+0.15}_{-0.14}$ & - & $0.29^{+0.15}_{-0.14}$ \\
        $\left(\bar{K}3\pi\right)^-$ & $0.38\pm 0.53$ & $0.38\pm 0.53$ & $0.38\pm 0.53$ \\
        $\left(\bar{K}4\pi\right)^-$ & $0.17\pm 0.37$ & $0.17\pm 0.37$ & $0.17\pm 0.37$ \\
        $\left(\bar{K}5\pi\right)^-$ & $0.03\pm 0.10$ & $0.03\pm 0.10$ & $0.03\pm 0.10$ \\
        \hline
        Sum & $15.86^{+1.60}_{-1.37}$ & $12.79^{+1.43}_{-1.66}$ & $28.65\pm 1.17$ \\
        \hline
    \end{tabular}
    \caption{Branching ratios for vector and axial vector current contributions to the strange sector of $\tau$ decays \cite{ALEPH:1999uux}. }
    \label{fig:VA_Separation}
\end{table}

\section{Details of the 2 Higgs doublet model}
\label{app:2hdm}

One of our benchmark models is a two-Higgs-doublet model, presented in~\cite{Kahn:2016vjr}, which contains non-neligible axial coupling. 
Here we provide some additional details to what is written in the main text.  
The two Higgs doublets, $H_u$ and $H_d$, are charged under a new $U(1)_D$ that induces both vector and axial couplings. 
In this model, the same Higgs doublet couples to all three generations of each type of fermion, implying that the $U(1)_{D}$ axial couplings are the same for each generation. 
In this way, we have two independent axial couplings, parameterized by the two Higgs charges, $q_{H_u}$ and $q_{H_d}$. 

After electroweak symmetry breaking, both Higgs doublets have acquired a vacuum expectation value, and we obtain mixing between the $X$ and the SM $Z$ boson. 
This defines the mixing angle $\theta_D\equiv \frac{2g_D}{g_z}\tilde{\theta}_D$ and the following charges:
\begin{align}
    \label{eq:2hdmchargeV}
    x^{u,c,t}_V &= \frac{1}{2}q_{H_u}  + \tilde{\theta}_D \left(\frac{1}{2} - \frac{4}{3} s_W^2 \right)\, , \quad
    x^{d,s,b}_V = \frac{1}{2}q_{H_d} + \tilde{\theta}_D \left(-\frac{1}{2} + \frac{2}{3} s_W^2 \right) \, , \nonumber \\
    x^e_V &= \frac{1}{2}q_{H_d} + \tilde{\theta}_D \left(-\frac{1}{2} + 2s_W^2 \right) \, , \quad
    x^\mu_V = x^e_V + \kappa \, , \quad
    x^\tau_V = x^e_V - \kappa \, , \nonumber \\
    x^{\nu_e}_V &= \frac{1}{2}\tilde{\theta}_D \, , \quad
    x^{\nu_\mu}_V = \frac{1}{2}(\tilde{\theta}_D + \kappa) \, , \quad
    x^{\nu_\tau}_V = \frac{1}{2}(\tilde{\theta}_D - \kappa) \, , \\
    \label{eq:2hdmchargeA}
    x^{u,c,t}_A &= -\frac{1}{2}q_{H_u} + \frac{1}{2} \tilde{\theta}_D \, ,  \quad
    x^{d,s,b}_A = x^s_A = -\frac{1}{2}q_{H_d} - \frac{1}{2} \tilde{\theta}_D \, , \nonumber\\
    x^{e,\mu,\tau}_A &= -\frac{1}{2}q_{H_d} - \frac{1}{2} \tilde{\theta}_D \, , \nonumber\\
     x^{\nu_e}_A &= x^{\nu_e}_V \, , \quad x^{\nu_\mu}_A = x^{\nu_\mu}_V \, , \quad x^{\nu_\tau}_A = x^{\nu_\tau}_V \, .
\end{align}
As an example, for $q_{H_u}=2, q_{H_d}=0.1, \tilde{\theta}_D=0.1$ the charges are written explicitly in \cref{tab:model charges} and the recasting can be seen in \cref{fig:2hdmrecast}. 
Below the hadronic threshold, recast bounds are presented by \cite{Kahn:2016vjr}.
Our results overlap with these for the relevant region and we obtain bounds for higher masses. 


\bibliographystyle{JHEP}
\bibliography{bib}

\end{document}